\documentclass{article}
\usepackage{graphicx} 
\usepackage{amsmath, amssymb}
\usepackage{mathtools}
\usepackage{xfrac}
\usepackage[table,xcdraw]{xcolor}
\usepackage{todonotes}
\usepackage[ruled,vlined,linesnumbered,algosection]{algorithm2e}
\usepackage{bm}
\usepackage{caption}
\usepackage{subcaption}
\usepackage{tikz}
\usepackage{multirow}
\usepackage{url}

\graphicspath{{img/}}

\newcommand{\doublebrackets}[1]{[\![ #1 ]\!]}
\newcommand{\knowingthat}{\,|\,}
\newcommand{\bc}{\boldsymbol{c}}
\newcommand{\by}{\boldsymbol{y}}
\newcommand{\bx}{\boldsymbol{x}}

\newcommand{\ba}{\boldsymbol{a}}
\newcommand{\bX}{\boldsymbol{X}}
\newcommand{\bY}{\boldsymbol{Y}}

\newtheorem{theorem}{Theorem}[section]

\newtheorem{remark}[theorem]{Remark}

\newtheorem{example}[theorem]{Example}

\newcommand{\modk}[1]{\textcolor{black}{#1}\index {#1}}
\newcommand{\modt}[1]{\textcolor{black}{#1}\index {#1}}
\newcommand{\modkz}[1]{\textcolor{black}{#1}\index {#1}}
\newcommand{\modtt}[1]{\textcolor{black}{#1}\index {#1}}
\newcommand{\nmodtt}[1]{\textcolor{black}{#1}\index {#1}}

\newtheorem{system}{Chemical System}

\title{A Hybrid tau-leap for simulating chemical kinetics with applications to parameter estimation }
\author{Thomas Trigo Trindade and  Konstantinos C.  Zygalakis}
\date{\today}

\begin{document}

\maketitle

\begin{abstract}
We consider the problem of efficiently simulating stochastic models of chemical kinetics. \modt{The Gillespie Stochastic Simulation algorithm (SSA) is often used to simulate these models, however, in many scenarios of interest, the computational cost quickly becomes prohibitive. This is further \modkz{exacerbated} in the Bayesian inference context when estimating parameters of chemical models, as the intractability of the likelihood requires multiple simulations of the underlying system}. To deal with issues of computational complexity in this paper, we propose a novel Hybrid $\tau$-leap algorithm for simulating well-mixed chemical systems. In particular,  the algorithm uses $\tau$-leap when appropriate (high population densities),  and SSA when necessary (low population densities, when discrete effects become non-negligible). In the intermediate regime, a combination of the two methods, which leverages the properties of the underlying Poisson formulation, is employed. As illustrated through a number of numerical experiments the Hybrid $\tau$ offers significant computational savings when compared to SSA without however sacrificing the overall accuracy. This feature is particularly welcomed in the Bayesian inference context, as it allows for parameter estimation of stochastic chemical kinetics at reduced computational cost. 
\end{abstract}


\section{Introduction}

In the last few \modkz{decades}, there has been an increase in the interest in biochemical systems with a small number of interacting components, see for example the phage $\lambda$-lysis decision circuit \cite{armcro98}, circadian rhythms \cite{vilar2002mnr} as well as the cell cycle \cite{kbp09}. 
In the setting of low copy numbers of interacting components, the stochastic variations may constitute a crucial element in the description of the dynamics of the systems, often in the form of bursts and cascading mechanisms that are typically not well captured by macroscopic models. 
Additionally, the general consensus now is that accounting for the stochasticity plays a central role in the interpretation of experimental data originating from cell and molecular processes \cite{armc97,swain2002iec}.

Even when incorporating stochasticity in the modelling of biochemical systems one needs to decide on the assumptions that hold for the system in question. In particular, when the underlying system is not well-mixed the appropriate microscopic description involves \modkz{describing the dynamics} of each chemical molecule separately \cite{doi1976std,erban2009smr}. On the other hand, when the system is sufficiently well mixed the kinetics of each species are described by a continuous time discrete space Markov chain, and in this case, the corresponding master equation is known as the Chemical Master Equation (CME) \cite{hi08}.  Essentially, the CME is a (potentially infinite-dimensional) system of Ordinary Differential Equations (ODEs) that describes, at each point in time, the probability density of all the different possible states of the system. 

Except for some very simple chemical systems \cite{jahu07}, due to the inherent high dimensionality of the CME, analytic solutions of the CME are not available. Therefore several methods have been developed \cite{WGM10,MB09,SCG17} that try to solve the corresponding system of differential equations directly. An alternative and more widely adopted approach relates to the direct simulation of the underlying Markov process. More precisely, the stochastic simulation algorithm (SSA) \cite{gillespie1977ess} exactly simulates trajectories whose probability density function matches that of the CME as the system evolves in time.   In addition, several alternative exact algorithms have been subsequently proposed \cite{gibr00,cao2004efs} that were shown to be computationally more efficient than SSA. The core idea behind these algorithms is that one samples a waiting time for the next reaction from an appropriate exponential distribution, while another draw of a random variable is then used to decide which of the possible reactions will occur.

A fundamental issue with all the exact algorithms described above is that running them can be computationally intensive for realistic problems. The reason behind this is that the time between subsequent reactions becomes very small leading thus to a computational bottleneck. {This issue is further \modkz{exacerbated} when one is interested 
in estimating parameters of the stochastic kinetics models from data, since the underlying likelihood is intractable, \modkz{\emph{i.e.,} not available in closed form, and one needs to perform multiple stochastic simulations, for example within a particle Markov chain Monte Carlo framework \cite{gowi11,wi20,shgogi14,APW20},  to deal with this intractability.}

One approach to deal with the computational complexity of exact algorithms like SSA, is to use an approximate algorithm such as $\tau$-leap \cite{gillespie2001aas}  in which the system is simulated over suitable time intervals for which several chemical events might occur. This lumping of chemical events can lead to significant computational savings \cite{hi08}. Furthermore, several variants of this algorithm have been proposed in the literature \cite{auger2006rla,TB04,tobu04,YB11}. An alternative approach to speeding up the SSA is to employ different approximations on the level of the description of the chemical system. A prime example of this is the reaction rate  \cite{hi08} equation (RRE). This is an ODE that is valid in the limit of large molecular populations, and it can be thought of as approximating the time evolution of the mean of the evolving Markov chain. An intermediate regime between the SSA and the reaction rate equation is the one where stochasticity is still important, but there exists a sufficient number of molecules to describe the evolving kinetics by a continuous model. This regime is called the chemical Langevin equation (CLE) \cite{gil00}, which is an \modkz{It\^{o}} stochastic differential equation (SDE) driven by a multidimensional Wiener process.

In practice, a lot of chemical systems can contain many different species with a wide range of population numbers. This multi-scale nature makes the direct application of approximate methods such as $\tau$-leap or of approximate models such as the CLE or the RRE non-trivial. This has motivated several different hybrid algorithms \cite{hepp2014ahs,SAFTA2015177,winkelmann2017hmc} that only treat certain chemical species as continuous variables and others as discrete.  By doing so, such schemes can benefit from the computational efficiency of continuum approximations (either deterministic or stochastic) while still taking into account discrete fluctuations when necessary. Such schemes typically involve partitioning the reactions into \emph{fast} and \emph{slow} reactions, with the fast reactions modeled using a continuum approximation (CLE or the reaction rate equation), while using the Markov jump process to simulate the discrete reactions. 
\modtt{As the effective firing rate of reactions depends on the present chemical populations, the reactions may effectively transition from displaying fast to slow behaviours if the involved chemical populations vary significantly over time.
This issue can be addressed by periodic re-partitioning \cite{hara02,radecr09} or by adopting a different approach based on \emph{population} scaling, i.e., adapting the type of reaction simulation as a function of the current population. Such an approach was followed in \cite{zydunerb16}, where an algorithm was proposed that would perform Langevin dynamics in regions of abundance, jump dynamics in regions where one of the involved chemical species is in small concentrations, and a mixture of both in intermediate regions.}

In this paper inspired by the work in \cite{zydunerb16} we propose a Hybrid $\tau$-leap scheme \modtt{(hereafter denoted Hybrid $\tau$)} that uses $\tau$-leaping dynamics to simulate reactions in which the discreteness cannot be discounted. \modkz{In particular, the proposed algorithm corresponds to a discretization of the chemical master equation and in the limit of small $\tau$ coincides with SSA. This is contrary to the algorithm proposed in \cite{zydunerb16}, where the limit of small timestep doesn't coincide with SSA. In addition, similar to the approach in \cite{zydunerb16}}   our scheme does not explicitly keep track of fast and slow reactions, but rather,  performs $\tau$-leap dynamics in regions of abundance, jump dynamics in regions where one of the involved chemical species is in small concentrations, and a mixture of both in intermediate regions. The preference of jump over $\tau$-leap dynamics is controlled for each individual reaction using a blending function which is chosen to take value $1$ in regions of low concentration, $0$ in regions where all involved chemical species are abundant, and smoothly interpolates in between. The choice of each blending region will depend on the reaction rate associated with the given reaction. The region should be generally chosen so that the resulting propensity is large in the $\tau-$ leaping region and small in the discrete region.

The rest of the paper is organized as follows. In Section \ref{sec:pre}, we review the standard approaches for simulating chemical kinetics such as SSA and the $\tau$-leap method. Furthermore, we introduce some basic ideas associated with parameter estimation for chemical kinetics, highlighting the fact that since the underlying likelihood is intractable one needs to design inference algorithms based on using fast and accurate simulations of the underlying chemical system. Then, in Section \ref{sec:main}, we introduce our new Hybrid $\tau$ algorithm, while in Section \ref{sec:numer} we perform a number of numerical simulations that demonstrate the excellent performance of the proposed numerical scheme when compared to other state-of-the-art methods. {We conclude in Section \ref{sec:concl} with a summary of our findings and a discussion of future directions.}  

\section{Preliminaries}
\label{sec:pre}
We will consider a biochemical network of $N$ species that interact through $M$ reaction channels within an isothermal reactor of fixed volume $V$. We will denote with $X_{i}(t), i=1,\ldots,N$ the number of molecules of species $S_{i}$ at time $t$ and let $\bX(t)=(X_{1}(t), \ldots,X_{N}(t))$. Throughout this work, we will assume that the chemical species are well mixed and hence   $\bX(t)$ can be modeled as a continuous time discrete space Markov process \cite{DG92}. More precisely, when in state $\bX(t)$, the $\modtt{j}$-th reaction gives rise to a transition $\bX(t) \rightarrow \bX(t)+\bm{\nu}_{j}$ with exponential distributed waiting time  with inhomogeneous rate $a_{j}(\bX(t))$ where $a_{j}(\cdot)$ and $\bm{\nu}_{j} \in \mathbb{Z}^{N}$ denote the propensity and stoichiometric vector corresponding to the $j$-th reaction, respectively. 

\modtt{Each reaction is of the form 
\[
\mu_{j,{1}}S_{1}+ \mu_{j,{2}}S_{2}+\ldots \mu_{j,{N}}S_{N} \xrightarrow{\modkz{c_j}} \mu_{j,{1}}'S_{1}+ \mu_{j,{2}}'S_{2}+\ldots \mu_{j,{N}}'S_{N}
\]
where $j=1,\ldots,M$ and $\mu_{j,{i}},\mu_{j,{i}}' \in \mathbb{N}=\{0,1,2,\ldots\}$, for $i=1,\ldots,N$. We will denote with $\bm{\mu}= (\mu_{j,{1}},\ldots,\mu_{j,{N}}), \bm{\mu}'= (\mu_{j,{1}}',\ldots,\mu_{j,{N}}')$ and then we have that the stoichiometric vectors $\bm{\nu}_{j}$, $j=1,\ldots,M$ satisfy
\[
\bm{\nu}_{j}= \bm{\mu}_{j}'-\bm{\mu}_{j}.
\]
}

These vectors describe how much the number of molecules change when the \modtt{$j$-th} reaction takes place. 
{For notational convenience, hereafter, $V = [\bm{\nu}_1, \ldots , \bm{\nu}_M]$.}
Under the assumptions of mass action kinetics, the associated propensity $a_{j}$ for the $j$-th reaction is
\[
a_{\modtt{j}}(\bx)=c_{\modtt{j}}\prod_{i=1}^{N}\frac{x_{i}!}{(x_{i}-\mu_{j,i})!}
\]
where $x_{i}$ is the number of molecules of $S_{i}$. 
{Again, for notational convenience, hereafter $\ba(\bx) = (a_1(\bx), \ldots, a_M(\bx))$.}

\subsection{Algorithms for simulating chemical kinetics}

The main assumption in modelling the evolution of $\bX(t)$ is that within the time interval $[t,t+dt)$, the probability of the reaction $j$ firing is proportional to $a_{j}(\bX(t))dt+o(dt)$. The process $\bX(t)$ can thus be expressed as the sum of $M$ Poisson processes with inhomogeneous rates $a_{j}(\bX(t))$. Furthermore, the process $\bX(t)$ can be expressed \cite{gil00,anku11} as a random time change of unit rate Poisson processes
\begin{equation}\label{eq:Pois}
\bX(t)=\bX(0)+\sum_{\modtt{j=1}}^{M}P_{\modtt{j}}\left(\int_{0}^{t}a_{\modtt{j}}(\bX(s))ds \right)\bm{\nu}_{\modtt{j}}    
\end{equation}
where $P_{\modtt{j}}$ are independent unit-rate Poisson processes. This formulation of the stochastic process is very helpful in terms of designing numerical methods that produce either exact or approximate samples. In particular, the standard way of sampling realisations of $\bX(t)$ is the Gillespie SSA \cite{gil77} see Algorithm \ref{alg:SSA}.

\begin{algorithm}[ht] 
 \caption{Stochastic Simulation Algorithm (SSA)} \label{alg:SSA}
 	\SetKwInOut{Input}{Input}
    \SetKwInOut{Output}{Output}
	\Input{$T > 0$, stoich. matrix $V$, propensities $\ba(\bx)$}
	\Output{Exact realisation of $\{\bX_t\}_{t \in [0,T]}$}
 $t = 0$\;
 \While{$t < T$}{
  Let $a_{0} =\sum_{j=1}^{M}a_{j}(\bX(t))$\;
  \nmodtt{Set} $\tau \nmodtt{=} -\log{u}/a_{0}$, where $u \sim \mathcal{U}[0,1]$ \;
 Choose the next reaction \modtt{$j$} with probability $a_{\nmodtt{j}}(\bX(t))/a_{0}$, where $\nmodtt{j}=1,\ldots,M$\;
  $\bX(t+\tau) = \bX(t) + \boldsymbol{\nu}_j $\;
  $t = t+\tau$. 
 }
\end{algorithm}
As we can see in Algorithm \ref{alg:SSA}  in order to advance the system from time $t$ to time $t+\tau$ one needs to generate two random variables. The next reaction method \cite{gibson2000ees} exploits further the structure of \eqref{eq:Pois} to provide a more efficient implementation of Gillespie's SSA when simulating systems with many reaction channels.  However, a fundamental computational issue with exact algorithms such as Gillespie's SSA or the next reaction method is they become computationally expensive when the number of molecules becomes large. In particular, in this case, the time to the next reaction becomes small, and if one is interested in simulating the chemical systems to time scales of $\mathcal{O}(1)$ will have to simulate a very high number of reaction events. 

One approach for speeding up exact algorithms is to further exploit the structure of \eqref{eq:Pois} to construct approximate algorithms. In particular, instead of explicitly calculating the time to the next reaction, one can choose a timescale of interest $\tau$ and then calculate how many reactions have occurred in each of the reaction channels. More precisely,  one can use the following approximation
\begin{equation} \label{eq:aprox_tau}
\int_{t}^{t+\tau}a_{j}(\bX(s))ds \simeq a_{j}(\bX(t))\tau
\end{equation}
and then using the formulation \eqref{eq:Pois}, \modkz{it} is not difficult to see that the number of reactions $k_{j}$ in the $j-$th channel can be approximated by 
\[
k_{j} \sim \mathcal{P}(\alpha_{j}(\bX(t))\tau),
\]
\modkz{where $P(\alpha_{j}(\bX(t))\tau)$ is a Poisson random variable with mean given by $\alpha_{j}(\bX(t))\tau$.}
The corresponding algorithm is called $\tau$-leaping \modkz{\cite{gillespie2001aas}}, see also Algorithm \ref{alg:tau-leap}. \modkz{A discussion about under which assumptions the approximation \eqref{eq:aprox_tau} is valid as well as strategies for choosing $\tau$ can be found in \cite{gil00,cagipe06, motevi14}. }

The main computational savings here come from the fact that several reaction events are lumped together, while in addition under appropriate assumptions on the propensity functions of the system \cite{cagipe06,anko12}  the error induced by this approximation is not very large. However, unlike exact methods like SSA in principle, the simple $\tau$-leap method might lead to negative populations, so one has to modify the original algorithm to avoid this issue \cite{TB04,cao2005anp}.

\begin{algorithm}[ht]
 \caption{$\tau$-leap}
 \label{alg:tau-leap}
 	\SetKwInOut{Input}{Input}
    \SetKwInOut{Output}{Output}
	\Input{$T > 0$, stoich. matrix $V$, propensities $\ba(\bx)$, $\tau > 0$}
	\Output{Approximate realisation of $\{\bX_t\}_{t \in [0,T]}$}
 $t\gets 0$\;
 \While{$t < T$}{
  \For{$j=1,\ldots,M$}{
  $k_j \sim \mathcal{P}(a_j(\bX(t)) \tau)$ \;
  }  
  $\bX(t+\tau) = \bX(t) + \sum_{j=1}^M \boldsymbol{\nu}_j k_j$\;
  $t = t+\tau$
 }
\end{algorithm}

\subsection{Parameter estimation for chemical kinetics} \label{subsec:paramestim}

\modk{In many practical applications one might be interested in estimating parameters of stochastic kinetics models such as reaction rates from time series data. In the typical setting \cite{andoho09,andoho10}  given a realisation of the stochastic process $\{\bX_t, t \in [0,T]\} \eqqcolon \bX_{[0,T]}$, the law of which depends on some parameter $\mathbf{c}$, we consider the problem of inferring the value of $\mathbf{c}$ while having access only to discrete and noisy observations \nmodtt{$\{\by_{i}, i=1, \ldots,L\}$} of $\bX_{[0,T]}$ \nmodtt{at discrete times $(t_1, \ldots,t_L) \subset [0,T]$.}}

\modk{In} our setting, the true data $\bx_{[0,T]}$ is a realisation of a stochastic chemical system $\bX_{[0, T]}$. That true data is (for simplicity) measured and saved at integer times $\doublebrackets{0,T} \coloneqq \{0, 1, \ldots, T\}$, during which measurement errors are possible. This leads to the noisy data $\by_{\doublebrackets{0,T}}$ (a realisation of $\bY_{\doublebrackets{0,T}}$). The observations $\bY_{\doublebrackets{0,T}}$ are assumed to be conditionally independent \nmodtt{among each other} given $\bX_{[0,T]}$. 

We aim to characterise or sample from the probability density $\mathbb{P}(\bc\knowingthat \by_{\doublebrackets{0,T}})$, to which end we use a Markov chain targetting that density. By Bayes' theorem,
\begin{align} \label{eq:markov-target}
    \mathbb{P}( \bc \knowingthat \by_{\doublebrackets{0,T}}) = \frac{\mathbb{P}(\by_{\doublebrackets{0,T}} \knowingthat \bc) \mathbb{P}_0(\bc)}
    {\mathbb{P}(\by_{\doublebrackets{0,T}})}
\end{align} 
Here, $\mathbb{P}_0(\bc)$ is \modkz{the prior assigned to $\bc$}. Using the law  of total probability on the denominator, we obtain 
\begin{equation} \label{eq:inta_likelihood}
    \mathbb{P}( \bc \knowingthat \by_{\doublebrackets{0,T}}) 
    \propto \mathbb{P}_0(\bc) \int \mathbb{P}(\by_{\doublebrackets{0,T}} \knowingthat \bx_{[0,T]} , \bc ) \mathbb{P}(\bx_{[0,T]} \knowingthat \bc)  d\bx_{[0,T]}.
\end{equation}
\modt{As the following example reveals, the likelihood $\modkz{\mathbb{P}(\by_{\doublebrackets{0,T}} \knowingthat  \bc )}$  is intractable, even in the case where there is no noise in the data $\by_{\doublebrackets{0,T}}$ and one observes $\bx_{[0,T]}$ directly.}
\begin{example} \label{ex:intra}
   We consider the simplest case where the data $\by_{[0,T]}$ coincide exactly with $\bx_{[0,T]}$. In this case, $\mathbb{P}(\by_{\doublebrackets{0,T}} \knowingthat  \bc)$  becomes
  \[\modkz{ 
   \mathbb{P}(\by_{\doublebrackets{0,T}}  \knowingthat \bc)=\prod_{i=1}^{T}p(\bX(t_{i}) \knowingthat \bc, \bX(t_{i-1}))},
   \]
   where $p(\bX(t_{i}) \knowingthat \bc, \bX(t_{i-1}))$ is the solution to the CME . However, except for some very simple chemical systems \cite{jahu07}, solutions to the CME are not analytically available which in turn implies that the likelihood $\mathbb{P}(\by_{\doublebrackets{0,T}}  \knowingthat \bc)$ is in general intractable.  
\end{example}

\modk{As the Example \ref{ex:intra} indicates the likelihood $\mathbb{P}(\by_{\doublebrackets{0,T}}  \knowingthat \bc)$ is intractable. There are different ways of dealing with this issue, one of which is through approximate Bayesian computation \cite{APW20}. However, here we choose to proceed by following the pseudo-marginal approach \cite{AR09}, similarly to what was done in \cite{shgogi14}. In particular, the idea is that if we have access to an unbiased estimator $\hat{\mathbb{P}}(\by_{\doublebrackets{0,T}}  \knowingthat \bc)$ of our intractable likelihood $\mathbb{P}(\by_{\doublebrackets{0,T}}  \knowingthat \bc)$ we can proceed in the standard manner to perform Bayesian inference within a Metropolis-Hastings framework by replacing the intractable likelihood by its unbiased estimator.}

\subsubsection{Particle Pseudo-Marginal \modkz{Metropolis–Hastings} algorithm}
\modk{As discussed above within the pseudo-marginal framework we need to have access to an unbiased estimator of our intractable likelihood. We do this by using a (bootstrap) particle filter \cite{smgo93} with importance resampling to iteratively construct the (unbiased) estimate of $\mathbb{P}(\mathbf{ \bc} \knowingthat \by_{\doublebrackets{0,T}})$. Combing this unbiased estimate with a Metropolis-Hastings step gives rise to the Particle Pseudo-Marginal Metropolis–Hastings algorithm (PPMMH)~\cite{andoho10}, see Algorithm \ref{alg:ppmmh}.}
\modt{In typical Metropolis-Hastings fashion, the state space is explored via a proposal kernel $q(\cdot \knowingthat \mathbf{c})$ generating proposals $\bc^{\star}$ from the current state $\bc$, and the proposals are kept in the chain using a Metropolis-Hastings accept/reject mechanism.}


\begin{algorithm}
    \caption{Particle Pseudo-Marginal Metropolis–Hastings }\label{alg:ppmmh}
    \SetKwInOut{Output}{Output}
    \Output{Samples $\bc_i \sim \mathbb{P}(\cdot | \by_{\doublebrackets{0,T}})$}
   \For{$i = 1, \ldots$}{ 
    generate proposal $\bc^{*} \sim q(\cdot \knowingthat \bc)$\; 
    compute $\hat{\mathbb{P}}(\by_{\doublebrackets{0,T}} \knowingthat \bc^*)$\;
    accept $\bc^{*}$ with probability  
        \begin{equation*}
          \min\left\{
           \frac{ 
           \hat{\mathbb{P}}(\by_{\doublebrackets{0,T}} \knowingthat \bc^{*}) \mathbb{P}_0(\bc^{*})}
           { \hat{\mathbb{P}}(\by_{\doublebrackets{0,T}}\knowingthat \bc^{(i-1)} ) 
           \mathbb{P}_0(\bc^{(i-1)})} 
           \times  
           \frac{q(\bc^{(i-1)} \knowingthat \bc^{*})}
           {q(\bc^{*} \knowingthat \bc^{(i-1)})} \;,\; 
        1
           \right\}.
        \end{equation*} 
  } 
\end{algorithm}

The boostrap particle filter (computed in line $2$ of~Algorithm~\ref{alg:ppmmh}) relies on the fact that, for $j>0$,
\begin{equation} \label{eq:xknowy}
    \mathbb{P}(\bx_{[0,j+1]} \knowingthat \by_{\doublebrackets{0,j+1}}) 
    \propto 
    \mathbb{P}(\by_{j+1} \knowingthat \bx_{j+1}) 
    \mathbb{P}( \bx_{[0,j]} \knowingthat \by_{\doublebrackets{0,j}}) 
    \mathbb{P}(\bx_{(j,j+1]} \knowingthat \bx_{[0, j]}).
\end{equation}
This allows refining the naive approach of simply simulating \modtt{$K$} realisations/particles up to time $T$. \modtt{To avoid ambiguities, we hereafter solely use the term ``particle'' to denote realisations of the particle filter, and not chemical species.}
The iterative method consists in propagating the particles over a length $1$ time interval, evaluating the likelihood of each particle given the data, and using an importance resampling mechanism. Among others, this allows to avoid the degeneracy of the filter~\cite{Doucet2000}. The unbiasedness of the estimator can be established using, e.g.,~\cite[p. 290]{andoho10}. \modtt{The initial step of the algorithm is performed by sampling from a given prior $\mathbb{P}_0(\bx_0)$.} 

The steps are summarised in Algorithm~\ref{alg:bpf}. 

\begin{algorithm}
    \caption{Boostrap Particle Filter with Importance Resampling }\label{alg:bpf}
    \SetKwInOut{Output}{Output}
    \SetKwInOut{Input}{Input} 
    \Input{Proposal $\bc^{*}$, \nmodtt{data $\by_{\doublebrackets{0,T}}$}}
    \Output{$ \hat{\mathbb{P}}(\by_{\doublebrackets{0,T}} \knowingthat \bc^{*})$}
    \For{$k = 1, \ldots,K$ \tcp*{Initialisation}}{
    draw $\bx^k_0 \sim \mathbb{P}_0(\bx_{0})$ (prior)
    }
    \For{$j = 0,1, \ldots, T-1$}{
    \For{$k = 1, \ldots, K$ \tcp*{Particles}}{    simulate $\bx_{(j,j+1]}^k \sim \mathbb{P}(\cdot \knowingthat \bx^k_{j} ,\, \bc^{*})$  \; \label{alglin:simulate}
    $(\bx^k_{[0,j+1)}, \tilde{\bx}^{k}_{j+1}) \gets (\bx^k_{[0,j]}, \bx^k_{(j,j+1]})$ \; 
    $w_{jk} = \mathbb{P}(\by_{j+1} \knowingthat \tilde{\bx}_{j+1}^{k})$ \;
    } 
    \For{$k = 1, \ldots, N$ \tcp*{Resampling}}{
    sample 
    $\bx^{k}_{j+1}$ from $\{\tilde{\bx}^{\ell}_{j+1}\}_{\ell=1}^N$ with resp. probability 
    $ 
        \frac{w_{j\ell}}{ \sum_{s=1}^N  w_{js}}.
    $
    } 
    }
    $
    \hat{\mathbb{P}}(\by_{\doublebrackets{0,T}} \knowingthat \bc^{*}) \gets \hat{\mathbb{P}}(\by_{0}) \prod_{j=1}^{T}  \left( \frac{1}{N}\sum_{k = 1}^N w_{jk} \right).
    $
\end{algorithm}


Line~\ref{alglin:simulate} of Algorithm~\ref{alg:bpf} entails the repeated simulation of $K$ particles over a time interval, resulting in a computationally intensive process. In this work, we speed up those computations by using the Hybrid $\tau$ algorithm. 

\modtt{\begin{remark}
Practically running the Bootstrap Particle Filter requires to make a few \nmodtt{algorithmic} choices and perform some amount of fine-tuning, chief among which are determining the number of particles $K$ and the type and parameters of the proposal kernel $q(\cdot | \bc)$. 
In this work, we make use of a Gaussian proposal kernel, and proceed in a bootstrap fashion by performing a sequence of exploratory runs in order to determine those quantities.
Those exploratory runs allow to determine a first estimate of the mean $\bc_{\mathrm{pre}}$ and the covariance of the kernel $\hat{\mathbf{C}}$ (up to a tuning constant $\gamma$). Following~\cite{shthal15, pisigi12}, the number of particles $K$ is then chosen such that the variance in the log-posterior $\mathrm{Var}(\mathrm{log}\,\hat{\mathbb{P}}(\by\knowingthat \bc_{\mathrm{pre}}))$ is around $2$.  
The tuning constant $\gamma$ the kernel proposal is also determined in a similar bootstrap fashion, again following the approach outlined in~\cite{shthal15}, with the aim that the accept-reject ratio is close to $10\%$. 
The practical details of our implementation are discussed in Section~\ref{subsubsec:methodology}. 
\end{remark}}


\section{Hybrid $\tau$-leap}
\label{sec:main}
We now introduce our proposed algorithm. The idea here is similar to the one in \cite{zydunerb16}. In particular, we will introduce one \emph{blending} function for each reaction denoted by $\beta_{j}(\bx):\mathbb{R}^{N}\mapsto [0,1], \ j=1, \ldots,M$. One can then simply rewrite equation \eqref{eq:Pois} in the following way
\[
\bX(t)=\bX(0)+\sum_{\modtt{j}=1}^{M}P_{\modtt{j}}\left(\int_{0}^{t} \beta_{\modtt{j}}(\bX(s))a_{\modtt{j}}(\bX(s))ds+\int_{0}^{t}(1-\beta_{\modtt{j}}(\bX(s)))a_{\modtt{j}}(\bX(s))ds \right)\bm{\nu}_{\modtt{j}}     
\]
Using the property of Poisson processes it is now possible to rewrite the equation above in the following \modkz{manner} 
\begin{eqnarray} \label{eq:Pois1}
\bX(t)   &=& \modtt{\bX(0)} +  \sum_{\modtt{j}=1}^{M}P_{\modtt{j}}\left(\int_{0}^{t} \beta_{\modtt{j}}(\bX(s))a_{\modtt{j}}(\bX(s))ds \right)\bm{\nu}_{\modtt{j}} \nonumber\\
         &+&  \sum_{\modtt{j}=1}^{M}P_{\modtt{j}}\left(\int_{0}^{t}(1-\beta_{\modtt{j}}(\bX(s)))a_{\modtt{j}}(\bX(s))ds \right)\bm{\nu}_{\modtt{j}}
\end{eqnarray}
This rewriting might appear trivial at first sight, but it is essential in terms of explaining our algorithm given the form of the blending functions $\beta_{r}$. In particular, for a single-species system, a natural choice of blending function is the following piecewise  linear function
\[
\beta(x,I_{1},I_{2})=\begin{cases}
			1, & \text{if $x \leq I_{1}$}\\
            \frac{I_{2}-x}{I_{2}-I_{1}}, & \text{if $I_{1} \leq x \leq I_{2}$} \\
            0,  & \text{if $x \geq I_{2}$}\\
		 \end{cases},
\]
\modkz{where $0 \leq I_{1} < I_{2}$}.
Furthermore,  in the case of a chemical system with $N$ species, we can construct blending functions in the following way. Let $R_{\modtt{j}}$ be the \modkz{indices} of the chemical species involved in the $\modtt{j}$-th reaction (both as reactants and products of reaction), \modkz{i.e $R_{j}=\{i: \mu_{j,{i}}\neq 0  \ \text{or} \ \mu'_{j,{i}} \neq 0 \}$}. Then we can define $\beta_{\modtt{j}}(\bx), \ \modtt{j} =1,\ldots,M$ as follows
\begin{equation} \label{eq:blend}
\beta_{\modtt{j}}(\bx)=1-\prod_{n \in R_{\modtt{j}}}(1-\beta(x_{n},I^{n,j}_{1},I^{n,j}_{2})),    
\end{equation}
where $I_{1}^{n,j}<I_{2}^{n,j}$, are the boundaries for each individual chemical species. 
\modtt{The boundaries $I^{n,j}_1$ and $I^{n,j}_2$ are user-defined and depend on the problem at hand -- in the general case, they may depend on both the species and the reaction.
They allow to separate the state space into a region where the reaction is simulated with SSA dynamics, pure $\tau$-leap or transitory hybrid dynamics in-between. As such, they should be chosen in such a way that the region where the $\tau$-leaping is applied is a valid numerical approximation. 
Two natural simplifications of this approach are when the boundaries depend only on the species ($I^{n,j}_i \equiv I^{n}_i$ for $i=1,2$) and when the boundaries are independent of species and reaction ($I^{n,j}_i \equiv I_i$ for $i=1,2$). In our numerical experiments, we restrict ourselves to the latter case. 
}

\modtt{We now define the following sets
\begin{subequations}
\begin{eqnarray}
C^j_{\mathrm{SSA}}&=&\{\bx \in \mathbb{N}^N \;|\; \exists n \in R_{j}, \ x_{n} \leq I^{n}_{1} \}, \\
C^j_{\tau\mathrm{-leap}} &=&\{\bx \in \mathbb{N}^N \;|\; \forall n \in R_{j}, \ x_{n} \geq I^{n}_{2} \}, \\
C^j_{\mathrm{hybrid}}&=& \mathbb{N}^N \setminus \left( C^j_{\mathrm{SSA}} \cup C^j_{\mathrm{\tau-{\text{leap}}}} \right).
\end{eqnarray}
\end{subequations}
}
It is not difficult to see that  
$\bx \in \cap_{\modtt{j}=1}^{M} C^{\modtt{j}}_{\mathrm{SSA}}$ translates to $\min{\beta_j(\bx)} = 1$, while  
$\bx \in \cap_{\modtt{j}=1}^{M} C^{\modtt{j}}_{\tau\mathrm{-leap}}$  corresponds when $\max{\beta_j(\bx)} = 0$. Hence combining \eqref{eq:blend} with \eqref{eq:Pois1} when $\bx \in \cap_{\modtt{j}=1}^{M} C^{\modtt{j}}_{\text{SSA}}$, we will use SSA to simulate \eqref{eq:Pois1}, while when $\bx \in \cap_{\modtt{j}=1}^{M} C^{\modtt{j}}_{\text{$\tau$-leap}}$ we will use the  $\tau$-leap to simulate \eqref{eq:Pois1}.  It is only in the region \modkz{ $\bx \in \cup_{\modtt{j}=1}^{M} C^{\modtt{j}}_{\mathrm{hybrid}}$ that some reactions have blending functions obtaining values in $(0,1)$ and it is for these reactions that we use a combination of SSA for the term $P_{\modtt{j}}\left(\int_{0}^{t} \beta_{\modtt{j}}(\bX(s))a_{\modtt{j}}(\bX(s))ds \right)$, and $\tau$-leap for the term $P_{\modtt{j}}\left(\int_{0}^{t} (1-\beta_{\modtt{j}}(\bX(s)))a_{\modtt{j}}(\bX(s))ds \right)$}. 
\nmodtt{Note that, except for the region where the system is simulated solely via SSA or $\tau$-leap, the simulation regime might be quite diverse, with some reactions being simulated with SSA, while others are simulated with $\tau$-leap and others still are in the hybrid regime. For convenience, we call this region ``mixed'' in Figure~\ref{fig:all_in_one}d, which is simply given by 
$$\mathbb{N}^{N} \setminus \left( \cap_{j=1}^R C^j_{\mathrm{SSA}} \cup \cap_{j=1}^R C^j_{\tau-\mathrm{leap}} \right).$$
} 

We call the resulting algorithm the Hybrid $\tau$-method  (see Algorithm \ref{alg:ha1-version2}). \modtt{Note that, for clarity, we denote by $\tau$-leap($\tau$, $\ba$) the shorthand for performing one $\tau$-leap step through Algorithm~\ref{alg:tau-leap}, only expressing its dependence on the time-step $\tau$ and the involved propensities $\ba$. Furthermore, for the propensities, the expression $\boldsymbol{\gamma} \ba$ is to be understood component-wise. 
}

\begin{algorithm}[ht]
 \caption{Hybrid $\tau$} 
  \label{alg:ha1-version2}
 	\SetKwInOut{Input}{Input}
    \SetKwInOut{Output}{Output}
\Input{$T > 0$, stoich. matrix $V$, propensities $\ba(\bx)$, blending functions $\{\beta_i\}_{i=1}^M$, $\delta t, \Delta t$}
\Output{Approximate realisation of $\{\bX_t\}_{t \in [0,T]}$} 
$t = 0$\;
 \While{$t < T$}{
  \uIf{\modtt{$\max_j{\beta_j(\bX_t) = 0}$}}{
  \modtt{$\bX(t+\Delta t)$ = $\tau$-leap($\Delta t$, $\ba(\bX_t)$)}  \;
  $t = t+\Delta t$ \;}
  \uElseIf{\modtt{$\min_j{\beta_j(\bX_t) = 1}$}}{
  \modtt{$a_0 = \sum_{j=1}^M a_j(\bX_t)$} \;
  $\tau = -\log(\xi_1)/a_{0}$, where $\xi_1 \sim \mathcal{U}(0,1)$ \;
  \modtt{Choose reaction $\bar{\jmath}$ with probability $a_{\bar{\jmath}}(\bX_t)/a_0$} \;
  \modtt{$\bX(t+\tau) = \bX_t + \boldsymbol{\nu}_{\bar{\jmath}}$} \;
  \modtt{$t = t + \tau$} \;
  }
  \uElse{
  \modtt{$a_0 = \sum_{j=1}^M \beta_j(\bX_t) a_j(\bX_t)$} \;
  $\tau = -\log(\xi_1)/a_{0}$, where $\xi_1 \sim \mathcal{U}(0,1)$ \;
\uIf{$\tau < \delta t$}{
  \nmodtt{Choose reaction $\bar{\jmath}$ with probability $\beta_{\bar{\jmath}}(\bX_t)a_{\bar{\jmath}}(\bX_t)a_0^{-1}$} \;
    \modtt{$\bX(t+ \tau)$ = $\tau$-leap($\tau$, $(1 - \boldsymbol{\beta}(\bX_t))\ba(\bX_t)$)}  \;
  \modtt{$\bX(t+\tau) = \bX(t+\tau) + \boldsymbol{\nu}_{\bar{\jmath}}$} \;
  $t = t+\tau$ \;}
  	 \uElse{ 
    \modtt{$\bX(t+ \delta t)$ = $\tau$-leap($\tau$, $(1 - \boldsymbol{\beta}(\bX_t))\ba(\bX_t$)}  \;
  $t = t+\delta t$. }
  }  
  }
\end{algorithm}

\begin{example} \label{ex:simple}
We consider  here the following chemical system 
\begin{align*}
S_{1} &\xrightarrow{c_1} \emptyset 
& \emptyset &\xrightarrow{c_2} S_{1}
 & S_{1}+S_{2} &\xrightarrow{c_3}  \emptyset 
 &\emptyset \xrightarrow{c_4}& S_{2}
\end{align*}
Figures \ref{fig:all_in_one}(a)-(c) illustrate the sets $C^{\modtt{j}}_{\mathrm{SSA}}, C^{\modtt{j}}_{\tau\mathrm{-leap}},C^{\modtt{j}}_{\nmodtt{\mathrm{hybrid}}}$ associated  with each of the reactions, here $\modtt{j}=1,\ldots,4$. In addition, in Figure \ref{fig:all_in_one}(d) we can see the partition of the state space in terms of which simulation regime is applied where. 
\end{example}

\begin{remark}
In Algorithm \ref{alg:ha1-version2} two different time-stepping strategies are being used. There is  $\delta t$ that relates to the time-step used by the $\tau$-leap method in the intermediate regime and  $\Delta t$ that relates to the time-step used by the $\tau$-leap method when only $\tau$-leap is used for simulation. This is done to provide extra flexibility but is not crucial for the performance of the algorithm. 
\end{remark}

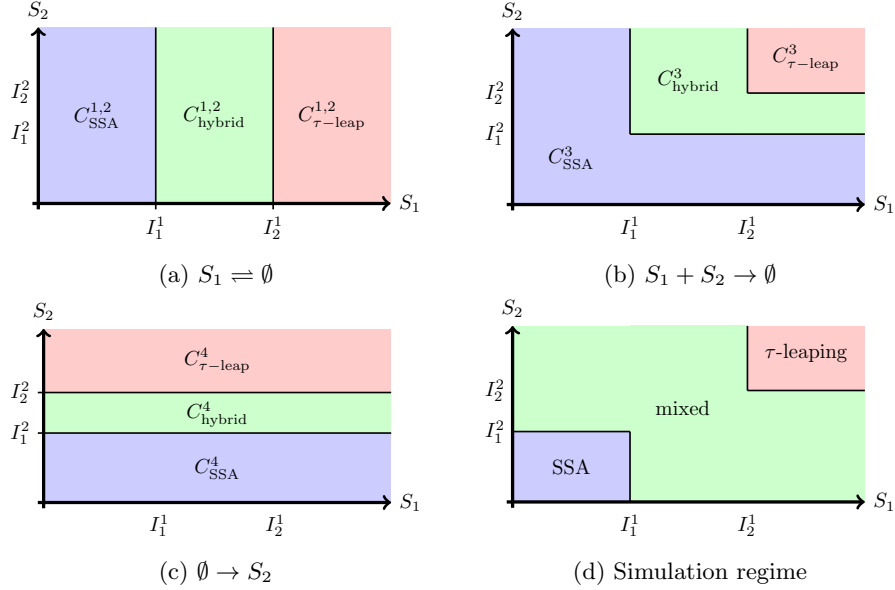
\begin{figure}
     \centering
     \begin{subfigure}[b]{0.48\textwidth}
         \centering
         \resizebox{\linewidth}{!}{
\begin{tikzpicture}
\fill[blue!20!white] (0,0) rectangle (2,3);
\fill[green!20!white] (2,0) rectangle (4,3);
\fill[red!20!white] (4,0) rectangle (6,3);
\draw[->,ultra thick] (-.1,0)--(6,0) node[right]{$S_1$};
\draw[->,ultra thick] (0,-.1)--(0,3) node[above]{$S_2$};
\draw[-,thick] (2,3)--(2,-.1) node[below]{$I^1_1$};
\draw[-,thick] (4,3)--(4,-.1) node[below]{$I^1_2$};
\node at (-.3, 1.2) {$I^2_1$} ; 
\node at (-.3, 1.9) {$I^2_2$} ;
\node at (1, 1.5) {$C^{1,2}_{\mathrm{SSA}}$} ;
\node at (3, 1.5) {$C^{1,2}_{\mathrm{hybrid}}$} ;
\node at (5, 1.5) {$C^{1,2}_{\tau\mathrm{-leap}}$} ;
\end{tikzpicture}
}        
         \caption{$S_1 \rightleftharpoons \emptyset$}
\end{subfigure}
     \hfill
     \begin{subfigure}[b]{0.48\textwidth}
         \centering
        \resizebox{\linewidth}{!}{
\begin{tikzpicture}
\fill[blue!20!white] (0,0) rectangle (2,3);
\fill[blue!20!white] (1.9,0) rectangle (6,1.2);
\fill[green!20!white] (2,1.2) rectangle (4,3);
\fill[green!20!white] (3.9,1.2) rectangle (6,1.9);
\fill[red!20!white] (4,1.9) rectangle (6,3);
\draw[->,ultra thick] (-.1,0)--(6,0) node[right]{$S_1$};
\draw[->,ultra thick] (0,-.1)--(0,3) node[above]{$S_2$};
\draw[-,thick] (4,3)--(4,1.9) ;
\draw[-,thick] (2,3)--(2,1.2) ;
\draw[-,thick] (6,1.2) -- (2,1.2) ;
\draw[-,thick] (6,1.9) -- (4,1.9) ;
\node at (-.3, 1.2) {$I^2_1$} ; 
\node at (-.3, 1.9) {$I^2_2$} ;
\node at (2, -.4) {$I^1_1$} ; 
\node at (4, -.4) {$I^1_2$} ; 
\node at (1, .8) {$C^3_{\mathrm{SSA}}$} ;
\node at (3, 2.1) {$C^3_{\mathrm{hybrid}}$} ;
\node at (5, 2.5) {$C^3_{\tau\mathrm{-leap}}$} ;
\end{tikzpicture}
}
         \caption{$S_1 + S_2 \rightarrow \emptyset $}
     \end{subfigure}
     \hfill
     \begin{subfigure}[b]{0.48\textwidth}
         \centering
         \resizebox{\linewidth}{!}{
\begin{tikzpicture}
\fill[blue!20!white] (0,0) rectangle (6,1.2);
\fill[green!20!white] (0,1.2) rectangle (6,1.9);
\fill[red!20!white] (0,1.9) rectangle (6,3);
\draw[->,ultra thick] (-.1,0)--(6,0) node[right]{$S_1$};
\draw[->,ultra thick] (0,-.1)--(0,3) node[above]{$S_2$};
\draw[-,thick] (6,1.2) -- (-.1,1.2) node[left]{$I^2_1$};
\draw[-,thick] (6,1.9) -- (-.1,1.9) node[left]{$I^2_2$};
\node at (3, .6) {$C^4_{\mathrm{SSA}}$} ;
\node at (3, 1.53)  {$C^4_{\mathrm{hybrid}}$} ;
\node at (3, 2.45)  {$C^4_{\tau\mathrm{-leap}}$};

\node at (2, -.4) {$I^1_1$} ; 
\node at (4, -.4) {$I^1_2$} ; 
\end{tikzpicture}
}
         \caption{$\emptyset \rightarrow S_2 $}
     \end{subfigure}
     \hfill
     \begin{subfigure}[b]{0.48\textwidth}
         \centering
        \resizebox{\linewidth}{!}{
\begin{tikzpicture}

\fill[blue!20!white] (0,0) rectangle (2,1.2);
\fill[green!20!white] (0,1.2) rectangle (2.1,3);
\fill[green!20!white] (2,0) rectangle (4,3);
\fill[green!20!white] (3.9,0) rectangle (6,1.9);
\fill[red!20!white] (4,1.9) rectangle (6,3);

\draw[->,ultra thick] (-.1,0)--(6,0) node[right]{$S_1$};
\draw[->,ultra thick] (0,-.1)--(0,3) node[above]{$S_2$};
\draw[-,thick] (4,3)--(4,1.9) ;
\draw[-,thick] (2,1.2)--(2,0) ;
\draw[-,thick] (6,1.9) -- (4,1.9) ;
\draw[-,thick] (2,1.2) -- (0,1.2) ;

\node at (-.3, 1.2) {$I^2_1$} ; 
\node at (-.3, 1.9) {$I^2_2$} ;
\node at (2, -.4) {$I^1_1$} ; 
\node at (4, -.4) {$I^1_2$} ; 

\node at (1, .6) {SSA} ; 
\node at (2.9, 1.6) {mixed} ; 
\node at (5, 2.55) {$\tau$-leaping} ; 
\end{tikzpicture}
}
         \caption{Simulation regime}
     \end{subfigure}
        \caption{(a)-(c): Illustration of $C^{\modtt{j}}_{\mathrm{SSA}}$, \nmodtt{$C^{j}_{\mathrm{hybrid}}$} and $C^j_{\tau\mathrm{-leap}}$ associated with each of the reactions of the chemical system in Example \ref{ex:simple}; (d): Partition of state space as per applicable simulation regime}
        \label{fig:all_in_one}
\end{figure}

\begin{remark}
Choosing one blending function per reaction is a modeling choice that tries to fully exploit the multiscale nature (when present) of the chemical kinetics. A more conservative approach would be to define a single blending function for all reactions, \modtt{i.e. $\displaystyle \tilde{\beta}(\bx) = 1 - \prod_{i=1}^N \beta(x_i, I_1, I_2)$.
This would cause the simulation regime of any reaction to be determined by the current state of all chemical species present in the system. 
}
As long as this choice of blending function sensibly partitions the state space, \emph{i.e.} ensuring that the underlying stochastic process spends some time outside the SSA region,  it would still lead to an algorithm that is more efficient than SSA.   
\end{remark}

\begin{remark} The idea of partitioning the state space is rather general. In particular, one could replace $\tau$-leap with the numerical method of their choice and the only thing that would need to be considered is how to do the simulation in the region of space where the numerical method co-exists with $\tau$-leap. An example of this is the Hybrid CLE method proposed in \cite{zydunerb16} where instead of using $\tau$-leap one simulates the term $P_{r}\left(\int_{0}^{t}(1-\beta_{\modtt{j}}(\bX(s)))a_{\modtt{j}}(\bX(s))ds \right)$ in \eqref{eq:Pois1} by using the diffusion approximation. 
\end{remark}

\section{Numerical Investigations}
\label{sec:numer}
We now present several different numerical experiments to illustrate the robustness \modk{and the accuracy} of our proposed approach. In particular, in Section \ref{subsec:comparison} we study three different model chemical systems and compare the performance of Hybrid $\tau$ with other algorithms. Furthermore, in Section \ref{subsec:par} we study the performance of the Hybrid $\tau$ method when used as the stochastic simulator of choice for parameter estimation as described in Section \ref{subsec:paramestim}. 
\subsection{Comparison with other numerical methods}
\label{subsec:comparison}
\subsubsection{Lotka-Volterra System}

We begin by considering a stochastic version of the Lotka-Volterra system. It is a first example where the Hybrid $\tau$ algorithm captures the correct statistical behaviour, while the standard CLE approximation {(with reflective boundary conditions)} completely fails to do so.  
The system is defined as:

\begin{system} \label{chem-sys:lv} \textbf{Lotka-Volterra System}
\begin{align*}
A &\xrightarrow{ \modtt{c_1}} 2 A 
& A + B &\xrightarrow{ \modtt{c_2}} 2B
 & B &\xrightarrow{ \modtt{c_3} } \emptyset
\end{align*}
\end{system}
The molecules of $A$ and $B$ are in a predator-prey relationship, both populations oscillating between states of abundance and scarcity. As in~\cite{zydunerb16}, the reaction constants are set to $c_1 = 2$, $c_2 = 0.002$, $c_3 = 2$. In Figure~\ref{fig:lv-ensemble}, a histogram is generated using $10^3$ SSA realisations simulated until $T=5$, which corresponds to one period of the solution to the corresponding Reaction Rate Equations. The initial conditions are chosen to be $A(0) = 50, B(0) = 60$. The system clearly exhibits a multiscale behaviour, spending time in all possible configurations of scarcity and abundance for both species. 
\nmodtt{Furthermore, a peculiarity of the system is that, for some realisations, the chemical species $B$ reaches $0$; in consequence, the system reduces to $A \rightarrow 2A$, i.e., exponential growth of the chemical species $A$.}
The locally high copy numbers of species cause the simulation via SSA to become excessively slow \nmodtt{(especially so in the situation of exponential growth, as the generated time steps evolve as $\mathcal{O}(A^{-1})$ for rapidly growing $A$, causing the simulation cost to become prohibitively expensive)}; this calls for employing approximate but accelerated schemes. 
\modkz{Here we will consider two such schemes the first being the CLE with reflective boundary conditions to ensure the non-negativity of the chemical species, and the second one being the Hybrid CLE proposed in \cite{zydunerb16}.}


\begin{figure}
\centering
\includegraphics[scale=.4]{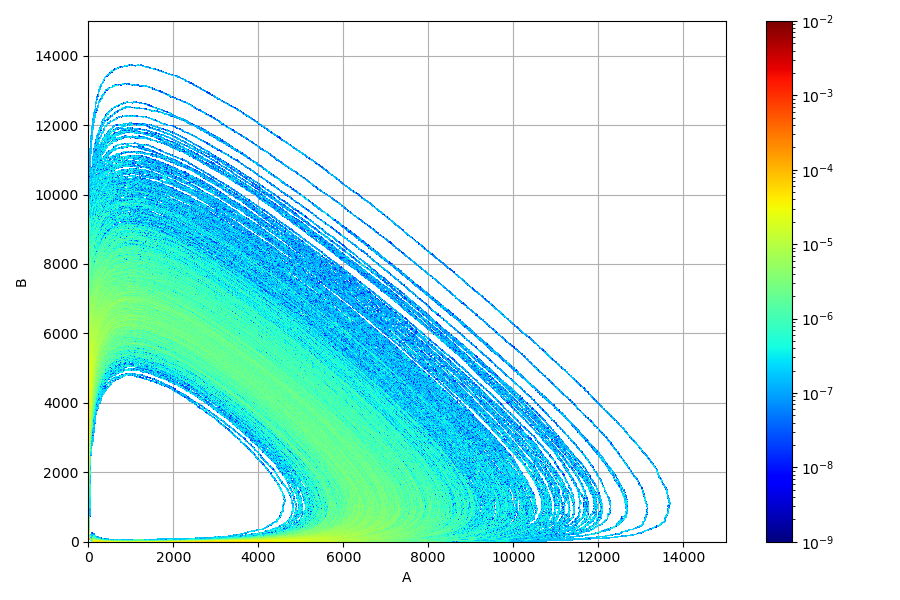}
\caption[Lotka-Volterra ensemble histogram]{Histogram of $10^3$ SSA simulations up to time $T=5$ of the Lotka-Volterra system with $A(0) = 50 , B(0) = 60$.  }
\label{fig:lv-ensemble}
\end{figure}

Figure~\ref{fig:lv-means-At} displays the numerical means of $A$ computed using respectively SSA, the Hybrid~$\tau$, the Hybrid CLE \cite{zydunerb16} and the CLE (with reflective boundary conditions) with $10^4$ samples. \modkz{We use an Euler-Maryama discretization for the CLE with timestep $\Delta t=10^{-2}$, while }
for the hybrid algorithms, the parameters are set to \modtt{$(I_1, I_2) = (5,10)$} and the step sizes $\Delta t = 10^{-2}$ and $\delta t = 10^{-3}$.  \modk{The step-sizes were chosen manually by computing the error for a number of short exploratory runs. A more sophisticated implementation of the Hybrid $\tau$ algorithm would require an adaptive scheme for the $\tau$-leap part of the process.} As we can observe, the numerical means of Hybrid $\tau$ and Hybrid CLE follow the same trend as SSA, whereas the CLE simulation completely fails to capture the right behaviour. 
\modtt{The mismatch is not due to the chosen stepsize in the CLE simulations, but rather to the fact that the solution to the CLE with reflective boundary conditions is fundamentally different from that of the Chemical Master Equation. This underlines the need for simulators directly based on the CME and not approximations thereof.}

\begin{figure}[h]
\centering
\begin{subfigure}{.5\textwidth}
  \centering
  \includegraphics[width=.9\linewidth]{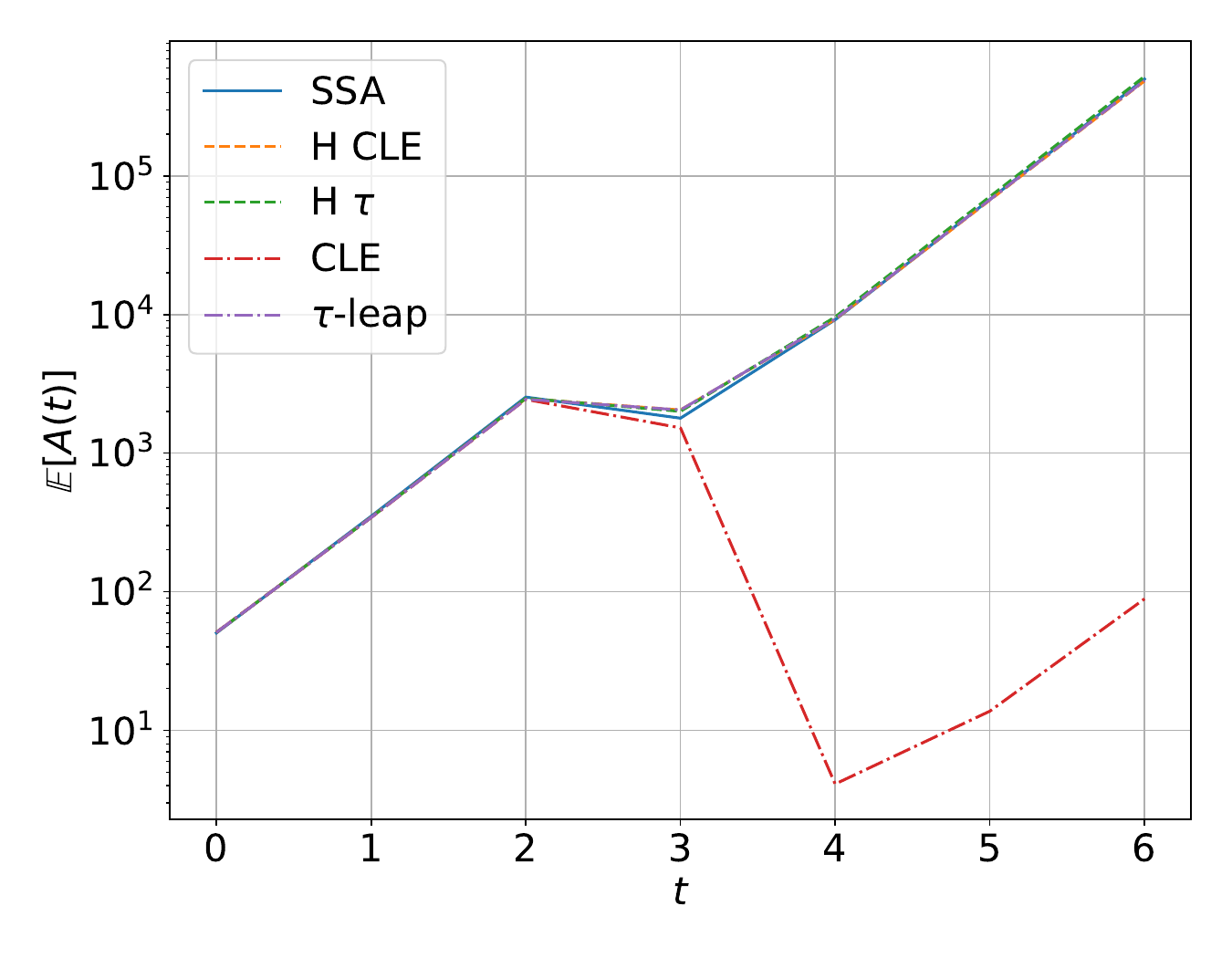}
  \caption{$\mathbb{E}[A(t)]$ for different simulators }
  \label{fig:lv-means-At}
\end{subfigure}%
\begin{subfigure}{.5\textwidth}
  \centering
  \includegraphics[width=.9\linewidth]{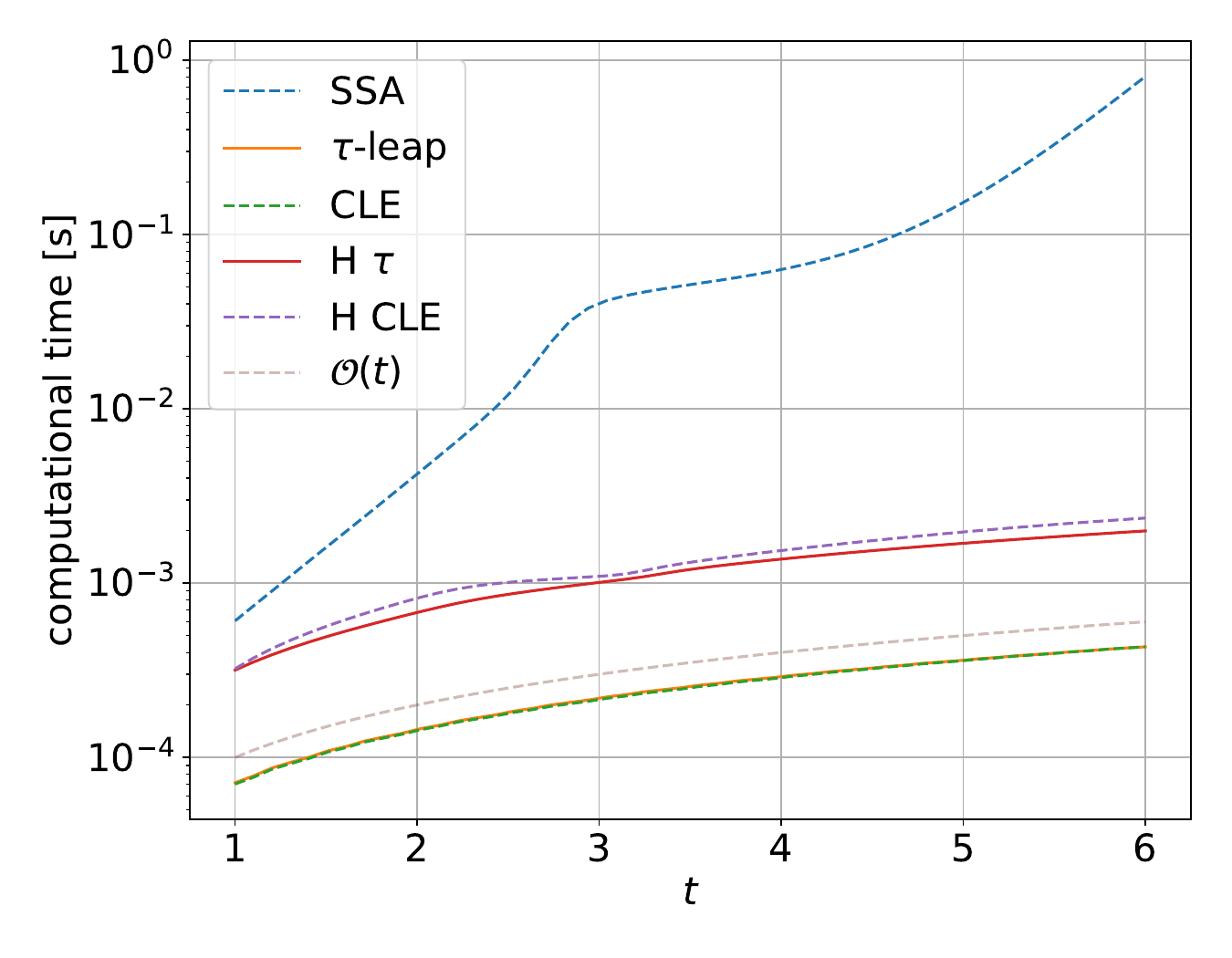}
  \caption{Average computational time }
  \label{fig:lv-timings}
\end{subfigure}
\caption{\nmodtt{Chemical System~\ref{chem-sys:lv} (Lotka-Volterra). Simulators : SSA, $\tau$-leap (with reflective boundary conditions), CLE (also with reflective BC), Hybrid CLE, Hybrid $\tau$. All averages were performed over $10^4$ simulations.}}
\label{fig:lv-system}
\end{figure}


\modtt{We now investigate the gains in computational efficiency. To this end, we simulate the Lotka-Volterra system up to time $T=6$ and measure the average computational time (using $10^4$ realisations) as a function of time, \nmodtt{using different simulation algorithms : SSA, $\tau$-leap, CLE, Hybrid CLE, and the Hybrid $\tau$. For the Hybrid algorithms, we used the blending region $(20,25)$, and the time steps are fixed to $\Delta t = \delta t = 10^{-2}$.}
The results are presented in Figure~\ref{fig:lv-timings}.
The computational time for the SSA grows exponentially for increasing $t$, while for the Hybrid $\tau$ simulations the computational cost grows linearly with time. At $T=6$, the computational cost of the SSA algorithm is \nmodtt{two} orders of magnitude higher than that of the Hybrid $\tau$ \nmodtt{and Hybrid CLE}. \nmodtt{Note that the Hybrid CLE and $\tau$-leap perform similarly well in this situation.} 
}


\subsubsection{\nmodtt{Birth-death system}}

\nmodtt{As a next example, we consider the following $1$-dimensional birth-death process
}
\begin{system} \label{chem-sys:birth-death}
\textbf{Birth-death system}
\begin{align*}
\mathrm{S} \xrightarrow{c_1} \varnothing &&
\mathrm{\varnothing} \xrightarrow{c_2} \mathrm{S}
\end{align*}
\end{system}
\nmodtt{
We modify the propensity function of the birth reaction to be $\tilde{a}_2(x) = c_2 \cdot \chi_{(0,N_{\mathrm{max}})}(x)$, which ensures the process never leaves the interval $[0,N_{\mathrm{max}}]$ \cite{zydunerb16}.  We are interested in calculating the amount of time that it takes for the system to reach zero when starting from the value $S = c_2 / c_1$. Note that due to our modification of the birth reaction, once the system reaches zero, it stays there, therefore we refer to this time as the extinction time.}  

\nmodtt{
In our experiments, we chose the parameters $N_{\mathrm{max}} = 50$, $c_1 = 1$ and $c_2 = 10$, and simulated the system over $10^4$ times to record extinction times via SSA, $\tau$-leaping, CLE, Hybrid CLE and Hybrid $\tau$. For the Hybrid $\tau$ and CLE, we used the blending region $(5,7)$ and time step $\Delta t = \delta t = 10^{-1}$. This is the timestep that $\tau$-leap and CLE use as well.  Figure~\ref{fig:met-qq} shows the quantile-quantile plot of extinction times obtained via $\tau$-leaping, CLE, Hybrid $\tau$ and Hybrid CLE, compared to those obtained from SSA. A quantile-quantile plot compares one set of quantiles from one distribution against another set. 
Therefore, the closer the plotted points are to the $y=x$ line, the more similar the distributions.
In that perspective, we see that $\tau$-leaping yields extinction time samples with a distribution significantly different to that of SSA, while the Hybrid $\tau$ matches the distribution much better; the same statement is true for CLE and Hybrid CLE. Figure~\ref{fig:met-cdfs} displays the same data, this time by comparing the empirical cumulative distribution functions. 
Furthermore, we report in Table~\ref{tab:birth-death} the relative speed-ups of each of the approximate algorithms with respect to SSA. As we can see all the approximate algorithms are faster than SSA, while the Hybrid $\tau$ is a bit faster than the Hybrid CLE while displaying similar accuracy with it. In principle further speeds ups could be observed for all the algorithms if one was to use larger time-steps but this would result is a loss in accuracy.    
}

\begin{table}[h]
\nmodtt{
\centering
\begin{tabular}{|l|l|l|}
\hline
\multicolumn{1}{|c|}{Algorithm} & \multicolumn{1}{c|}{Avg sim. time {[}s{]}} & \multicolumn{1}{c|}{Ratio SSA} \\ \hline
SSA   & 1.760409e-02 & 1       \\
H-tau & 1.258614e-02 & 1.39869 \\
Tau-L & 7.784003e-03 & 2.26157 \\
H-CLE & 1.689645e-02 & 1.04188 \\
CLE   & 8.669714e-03 & 2.03053 \\ \hline
\end{tabular}
\caption{\nmodtt{Simulation times of the Birth-Death system until extinction, averaged over $10^4$ simulations.}}
\label{tab:birth-death}
}
\end{table}

\begin{figure}[h]
\centering
\begin{subfigure}{.5\textwidth}
  \centering
  \includegraphics[width=.98\linewidth]{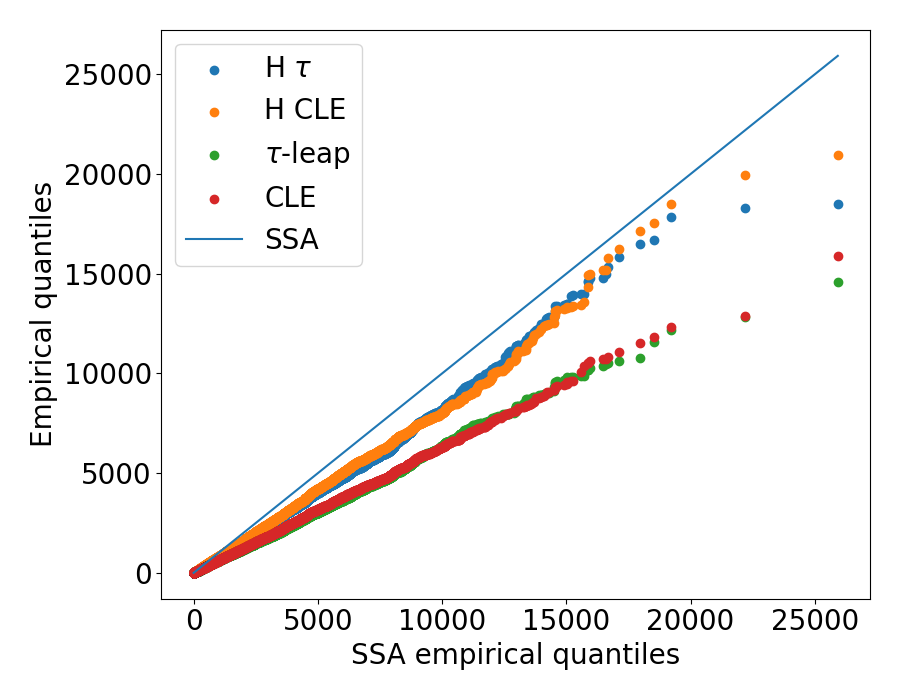}
  \caption{Quantile-quantile comparison}
  \label{fig:met-qq}
\end{subfigure}%
\begin{subfigure}{.5\textwidth}
  \centering
  \includegraphics[width=.98\linewidth]{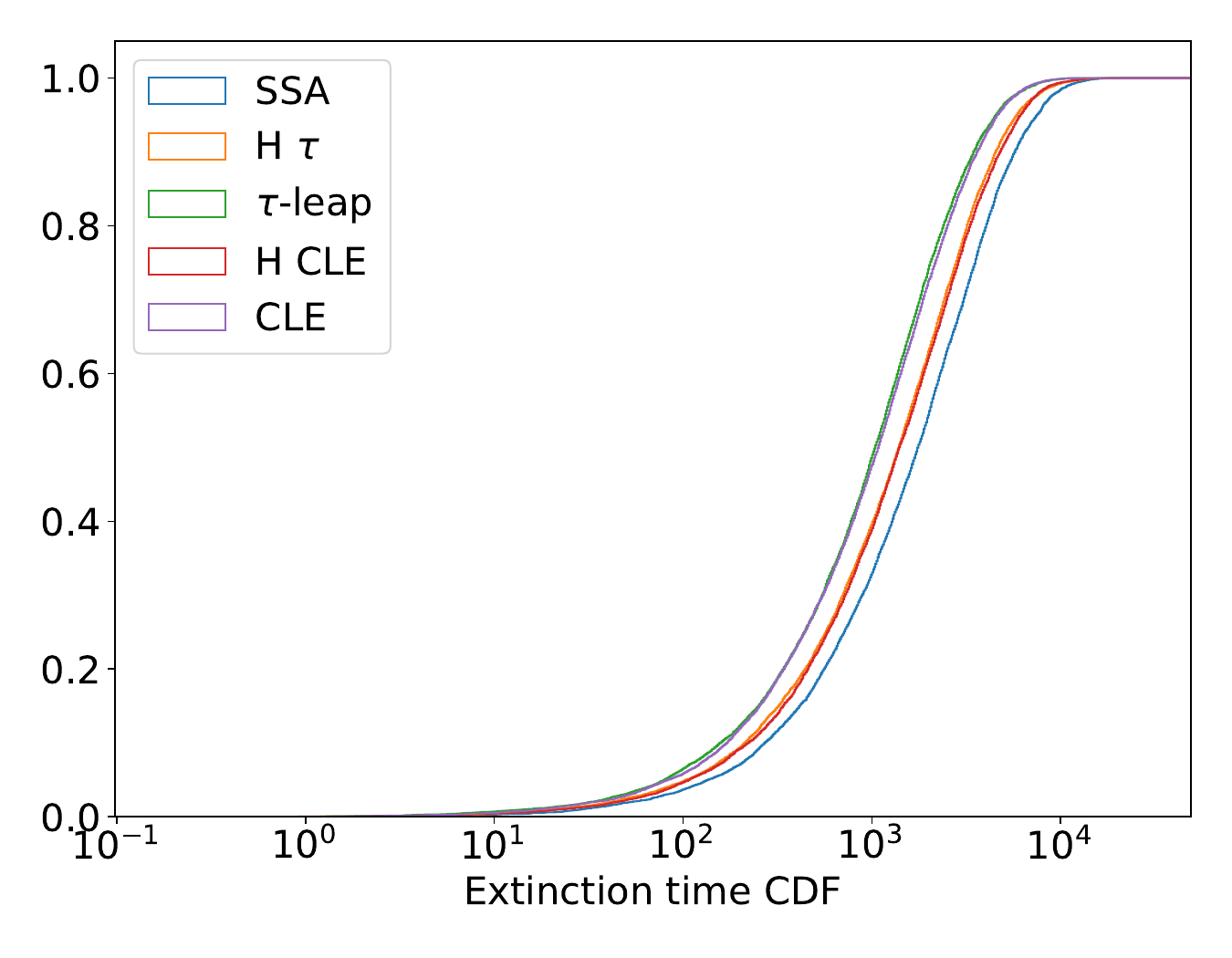}
  \caption{Cumulative CDFs comparison}
  \label{fig:met-cdfs}
\end{subfigure}
\caption{\nmodtt{Birth-death system. Statistical comparisons of extinction times obtained via SSA, $\tau$-leap, CLE, Hybrid $\tau$ and Hybrid CLE, taken over $10^4$ simulations.}}
\label{fig:met-comparisons}
\end{figure}

\subsubsection{Schl\"{o}gl System}

We now study the Schl\"{o}gl system taken from~\cite{abgaro21}. It is a non-linear model where the density function of the main reactant $S$ displays bistability for a certain choice of reaction constants. 
\begin{system} \label{chem-sys:schlogl}
\textbf{Schlögl System}
\begin{align*}
B_1 + 2S \xrightleftharpoons[c_2]{c_1} 3S && 
B_2 \xrightleftharpoons[c_4]{c_3} S,
\end{align*}
\end{system}
\noindent where $B_1$ and $B_2$ are buffered species, i.e. their populations are kept at constant values $N_1 = 10^5 $ and $N_2 = 2\cdot 10^5$ respectively. In effect, only $S$ needs to be tracked and this results in the following propensity functions: 
\begin{align*}
a_1(x) = \frac{c_1  }{2}N_1 x (x-1), && a_2(x) = \frac{c_2}{6}x(x-1)(x-2), && a_3(x) = c_3 N_2, && a_4(x) = c_4 x.
\end{align*}
We set $c_1 = 3 \cdot 10^{-7}$, $c_2 = 10^{-4}$, $c_3 = 10^{-3}$ and $c_4 = 3.5$ 
\modk{In Figure~\ref{fig:schlogl-long-path} we plot one trajectory of the system for this choice of parameters for time $T=10^{6}$ using the SSA. As we can see  the system exhibits a bistable behaviour as it tends to send time around two peaks one located around $80$ and one around $560$.}

We experimented \modk{similarly} to what was done in~\cite{abgaro21}, comparing the numerical probability distributions obtained from the SSA and Hybrid $\tau$. For \modtt{the Hybrid $\tau$, different combinations of blending regions $(I_1, I_2)$ and time steps $(\Delta t, \delta t)$ were tested.}
The initial condition is given by $S(0) = 250$. 
For each combination of parameters, $10^6$ paths were generated up to time $T=50$. \modk{Note that this time is not large enough so as for the system to jump between the two peaks of distribution as we can see in Figure \ref{fig:schlogl-paths} (b) in which we plot $10$ trajectories of the system using SSA.} 

\begin{figure}
\centering
\begin{subfigure}{.5\textwidth}
  \centering
  \includegraphics[width=.95\linewidth]{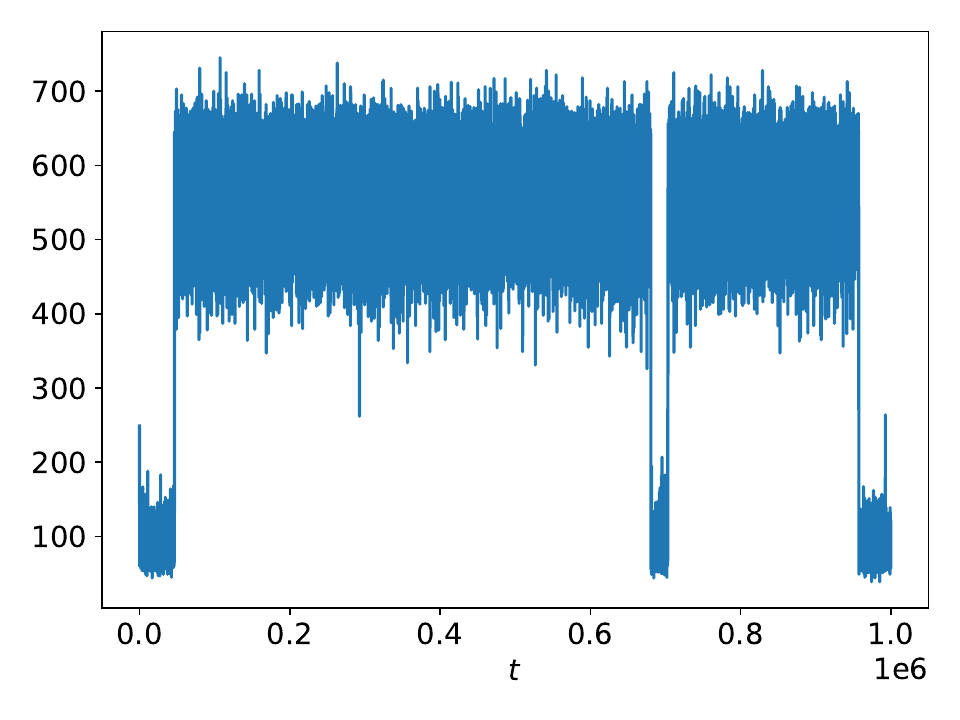}
  \caption{$T=10^{6}$}
  \label{fig:schlogl-long-path}
\end{subfigure}%
\begin{subfigure}{.5\textwidth}
  \centering
  \includegraphics[width=.95\linewidth]{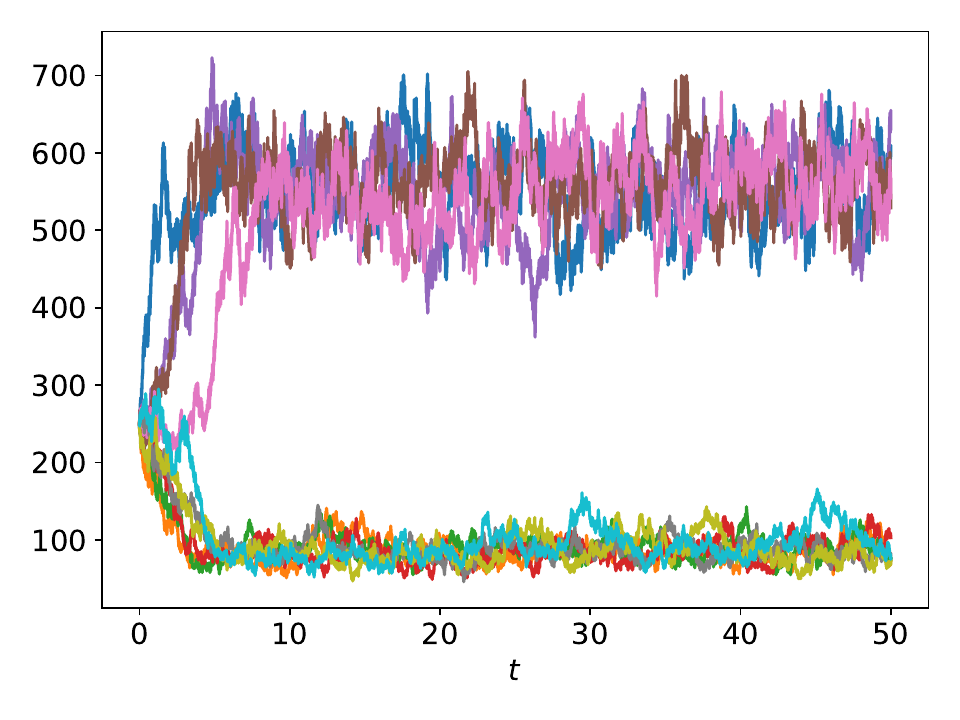}
  \caption{$T=50$ (10 trajectories)}
  \label{fig:sub2}
\end{subfigure}
\caption{Trajectories of the Schl\"{o}gl chemical system }
\label{fig:schlogl-paths}
\end{figure}

These results are presented in Figure~\ref{fig:schlogl-2}, in which the lower plot represents the numerical probability distributions, and the upper plots are zoom-ins corresponding to the intervals where the peaks are located. 
It is seen that, when using $\Delta t = 10^{-2}$ (parameter sets $1-4$), the position of the blending region and its width hardly affect the probability distribution, which aligns very well with that obtained via SSA.
In contrast, a numerical bias is clearly visible when using the time-step $\Delta t = 0.25$ (parameter sets $5-8$). In the high peak region (upper right zoom-in of Figure~\ref{fig:schlogl-2}), the obtained numerical densities concentrate on the same probability distribution which is not aligned with the SSA distribution. \nmodtt{This is of course to be expected, since the Hybrid $\tau$ method coincides with standard $\tau$-leap in this region due to the choice of the blending function.}

The numerical densities appear less stable in the first peak. 
The kink in the parameter set $5$ density is likely due to the combination of the sharply varying step size between $\delta t$ and $\Delta t$ (two orders of magnitude), the transition from the hybrid regime to the pure $\tau$-leaping regime at $S=80$, and the non-trivial interplay of dynamics oscillating between these two regions. 
This is further justified by the fact that the distribution associated to the parameter set $1$, which has the same blending function but smaller $\Delta t$, does not display such a behaviour and on the contrary is well aligned with the SSA density. 
}

\modtt{
Table~\ref{tab:schlogl-comparison-times} describes the average time taken (over $10^3$ realisations) for the simulation of realisations oscillating near the low peak, respectively the high peak. Note that for $T=50$ the probability of switching between these two peaks is very small indeed.
The last two columns display the ratio between the average time taken by SSA to simulate realisations in the low peak (resp. high peak) and that of the Hybrid $\tau$. 
This illustrates the potential computational benefits of the Hybrid $\tau$ but also the caveats of the algorithm. 
In all cases except for the simulation of the low peak when using $(I_1 I_2) = (50, 200)$, Hybrid $\tau$ outperforms SSA.   
The slow-down in the latter case can be attributed to the fact that a large portion of the mass of the probability density function (and hence, where the algorithm will spend more time) is located in a part that is still simulated in part with SSA with propensities weighted by $\beta$.  
Observing a speed-up when the system mainly evolves in the hybrid region is still possible, see the lines corresponding to parameters $(I_1, I_2) = (40, 100)$ in Table~\ref{tab:schlogl-comparison-times}.
This however is only possible when carefully choosing the time step in the hybrid dynamics region. 
Indeed, choosing $\delta t$ too big will cause the Hybrid $\tau$ to often accept the next reaction time proposed by the SSA process with modified propensities, hence causing the algorithm to advance (at most) at the speed of the SSA. 
On the other hand, choosing $\delta t$ too small causes the algorithm to simply advance at the speed of that small time step, which might be smaller than just propagating the system with SSA.  
A compromise is then found when the chosen time-step allows to consistently skip a few reactions within the hybrid dynamics region, i.e., it should correspond to a few multiples of the average time until next reaction of the SSA dynamics with propensities weighted by $\beta$. 
Furthermore, we remark that these speed-ups are obtained even with the computational overhead that the Hybrid $\tau$ carries at each iteration through its conditional logic.
}


\begin{table}
\centering
\begin{tabular}{ccllll|ll|ll|}
\cline{7-10}
\multicolumn{1}{l}{} &
  \multicolumn{1}{l}{} &
   &
   &
   &
   &
  \multicolumn{2}{c|}{Timings {[}s{]}} &
  \multicolumn{2}{c|}{Ratio SSA} \\ \cline{2-10} 
\multicolumn{1}{l|}{} &
  \multicolumn{1}{c|}{\begin{tabular}[c]{@{}c@{}}Param.\\ set\end{tabular}} &
  \multicolumn{1}{c|}{$\Delta t$} &
  \multicolumn{1}{c|}{$\delta t$} &
  \multicolumn{1}{c|}{$I_1$} &
  \multicolumn{1}{c|}{$I_2$} &
  \multicolumn{1}{c|}{LP} &
  \multicolumn{1}{c|}{HP} &
  \multicolumn{1}{c|}{LP} &
  \multicolumn{1}{c|}{HP} \\ \hline
\multicolumn{1}{|c|}{} &
  \multicolumn{1}{c|}{1} &
  1.0e-2 &
  \multicolumn{1}{l|}{5e-3} &
  40 &
  80 &
  \multicolumn{1}{c}{5.47e-03} &
  3.51e-03 &
  1.801 &
  38.05 \\
\multicolumn{1}{|c|}{} &
  \multicolumn{1}{c|}{2} &
  1.0e-2 &
  \multicolumn{1}{l|}{1e-2} &
  20 &
  40 &
  3.03e-03 &
  3.49e-03 &
  3.2531 &
  38.161 \\
\multicolumn{1}{|c|}{} &
  \multicolumn{1}{c|}{3} &
  1.0e-2 &
  \multicolumn{1}{l|}{2e-3} &
  50 &
  200 &
  2.98e-02 &
  3.51e-03 &
  0.3307 &
  38.008 \\
\multicolumn{1}{|c|}{} &
  \multicolumn{1}{c|}{4} &
  1.0e-2 &
  \multicolumn{1}{l|}{1e-2} &
  40 &
  100 &
  7.88e-03 &
  3.53e-03 &
  1.2512 &
  37.779 \\
\multicolumn{1}{|c|}{} &
  \multicolumn{1}{c|}{5} &
  2.5e-1 &
  \multicolumn{1}{l|}{5e-3} &
  40 &
  80 &
  2.77e-03 &
  1.38e-04 &
  3.554 &
  968.82 \\
\multicolumn{1}{|c|}{} &
  \multicolumn{1}{c|}{6} &
  2.5e-1 &
  \multicolumn{1}{l|}{1e-2} &
  20 &
  40 &
  1.42e-04 &
  1.34e-04 &
  69.506 &
  993.38 \\
\multicolumn{1}{|c|}{} &
  \multicolumn{1}{c|}{7} &
  2.5e-1 &
  \multicolumn{1}{l|}{2e-3} &
  50 &
  200 &
  2.95e-02 &
  1.39e-04 &
  0.33435 &
  957.39 \\
\multicolumn{1}{|c|}{\multirow{-8}{*}{H $\tau$}} &
  \multicolumn{1}{c|}{8} &
  2.5e-1 &
  \multicolumn{1}{l|}{1e-2} &
  40 &
  100 &
  6.60e-03 &
  1.36e-04 &
  1.4925 &
  983 \\ \hline
\multicolumn{1}{|c|}{SSA} &
  \cellcolor[HTML]{C0C0C0} &
  \cellcolor[HTML]{C0C0C0} &
  \cellcolor[HTML]{C0C0C0} &
  \cellcolor[HTML]{C0C0C0} &
  \cellcolor[HTML]{C0C0C0} &
  9.85e-03 &
  1.33e-01 &
  1 &
  1 \\ \hline
\end{tabular}
\caption{ \modtt{\nmodtt{Schl\"ogl system} simulation times for SSA and Hybrid $\tau$ (H $\tau$) with multiple parameters. LP=low peak, HP=high peak. Timings show the \nmodtt{average runtime (over $10^3$ simulations) to simulate a realisation in LP or HP over the time interval $[0,50]$}.
Ratio SSA columns show the ratio of that time with the time needed by the SSA for the same peak.}}
\label{tab:schlogl-comparison-times}
\end{table}

\begin{figure}[h]
\centering
\includegraphics[scale=.44]{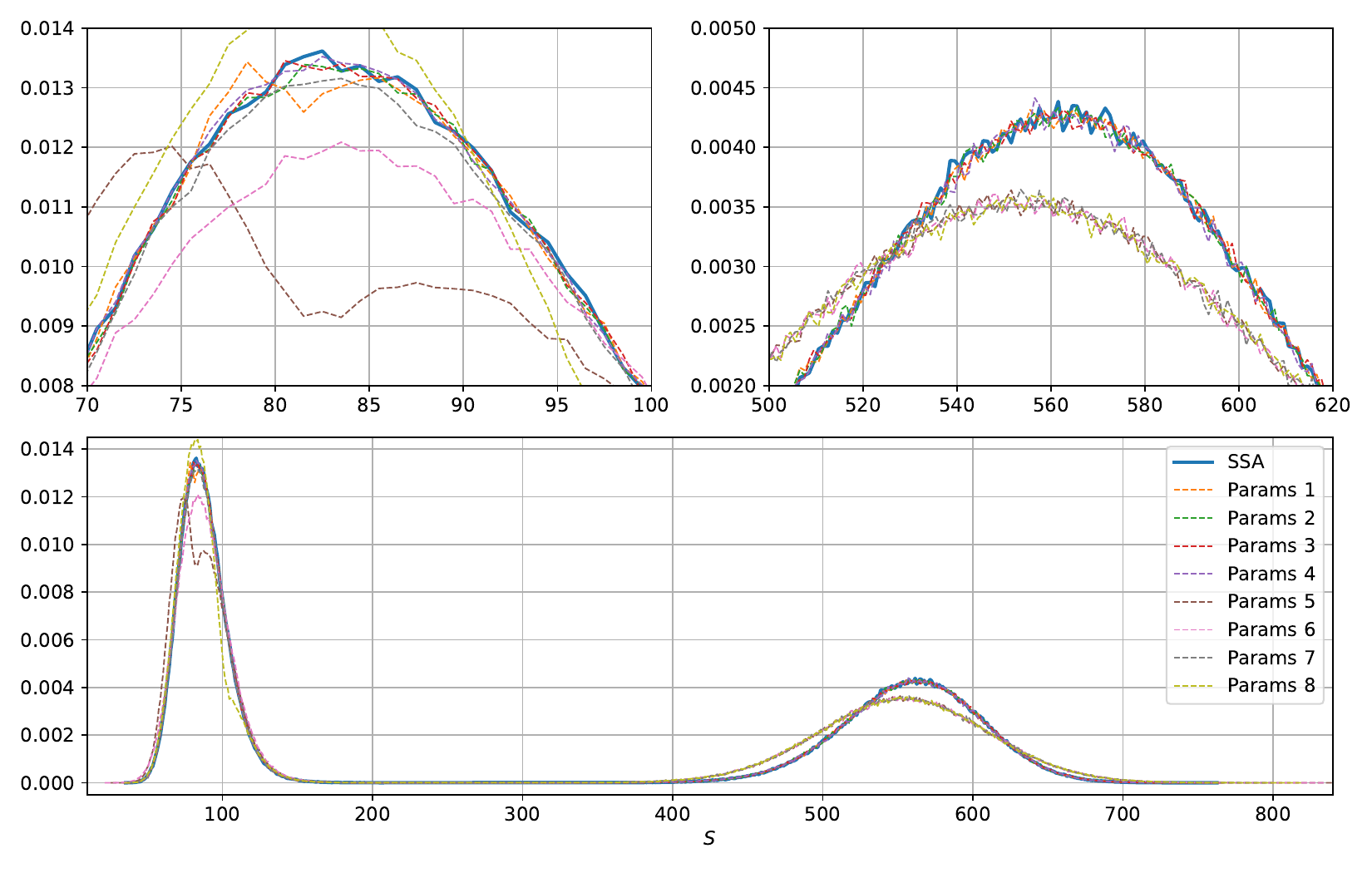}
\caption[Schlögl system density function through Hybrid $\tau$]{\modtt{Numerical pdfs at $T=50$ for the Schl\"ogl system using Hybrid $\tau$, for different parameter sets of $(\Delta t, \delta t, I_1, I_2)$. See Table~\ref{tab:schlogl-comparison-times} for the values of each parameter set.
Upper left plot : zoom-in of numerical pdfs on $[70,100]$. Upper right plot : zoom-in on $[500, 620]$. }
}
\label{fig:schlogl-2}
\end{figure}


\subsubsection{Autorepressive System}

Finally, we consider the autorepressive genetic system, taken from~\cite{schwi17}. This is yet another example where chemical reactions happen on different scales. In this gene regulatory system, the DNA is only present in one or two copies (active or inactive), the messenger RNA (mRNA) and proteins 
may reach much higher numbers. The system is \modk{described} as follows: 
\begin{system} \label{chem-sys:ags}
\textbf{Autorepressive Gene System}
\begin{align*}
\mathrm{DNA} \xrightarrow{c_1} \mathrm{DNA} + \mathrm{mRNA} && \mathrm{mRNA} \xrightarrow{c_2} \mathrm{mRNA} + \mathrm{P} && \mathrm{DNA} + \mathrm{P} \xrightarrow{c_3} \mathrm{DNA}_0 \\
\mathrm{mRNA} \xrightarrow{c_5} \emptyset && \mathrm{P} \xrightarrow{c_6} \emptyset && \mathrm{DNA}_0  \xrightarrow{c_4} \mathrm{DNA} + \mathrm{P}
\end{align*}
\end{system}
In essence, the active DNA is transcribed to mRNA at rate $c_1$, which in turn produces proteins (P) at rate $c_2$. A protein may repress the active DNA, causing it to become inactive (represented by DNA$_0$) at rate $c_3$. The inactive DNA reactivates at rate $c_4$. Finally, the mRNA and the proteins are degraded at rates $c_5$ and $c_6$ respectively. 

The reaction rates are set to $c_1 = 10^{-2}$, $c_2 = 0.5$, $c_3 = 0.1$, $c_4 = 10^{-2}$, $c_5 = 5 \cdot 10^{-3}$ and $c_6 = 0.2$. \modkz{Figure 7, shows a simulation of the system up to time $T=10^{3}$ using SSA. As we can see DNA and DNA$_{0}$ obtain very small values but mRNA and P obtain quite a long range of values. }

\modkz{We now simulate this chemical system with the Hybrid $\tau$ \nmodtt{and the Hybrid CLE} for three different blending regions namely  $(5,10), (10,15)$ and $(15, 20)$}; the time steps were set to $\Delta t = 10^{-2}$ and $\delta t = 10^{-3}$. For each combination of parameters, $10^4$ simulations were performed up to time $T = 10^3$. The means and variances of these different processes are detailed in Tables~\ref{tab:ags-means} and~\ref{tab:ags-stds} \nmodtt{(in Appendix)}. 
The results \nmodtt{are again fairly acccurate for all combinations of parameters and algorithms}, and indicate a certain robustness with respect to the chosen blending region.
\nmodtt{Albeit somewhat marginal, the Hybrid $\tau$ seems to perform slightly more accurately than the Hybrid CLE  when using bigger time steps.}
\nmodtt{We do not compare the efficiency with respect to $\tau$-leap or CLE; simulation via either of these methods would immediately break the ``switch'' nature of the DNA, which is either activated ($\mathrm{DNA} = 1$ and $\mathrm{DNA}_0 = 0$) or not ($\mathrm{DNA} = 0$ and $\mathrm{DNA}_0 = 1$).}
\nmodtt{Furthermore, in Table~\ref{tab:ags-times} in Appendix, we report the times taken by each algorithm for each combination of parameters, as well as their speed-up compared to SSA. Both methods display a modest but consistent speed-up with respect to SSA, up to $3$ times faster.}

\begin{figure}[h]
    \centering
    \includegraphics[scale=.4]{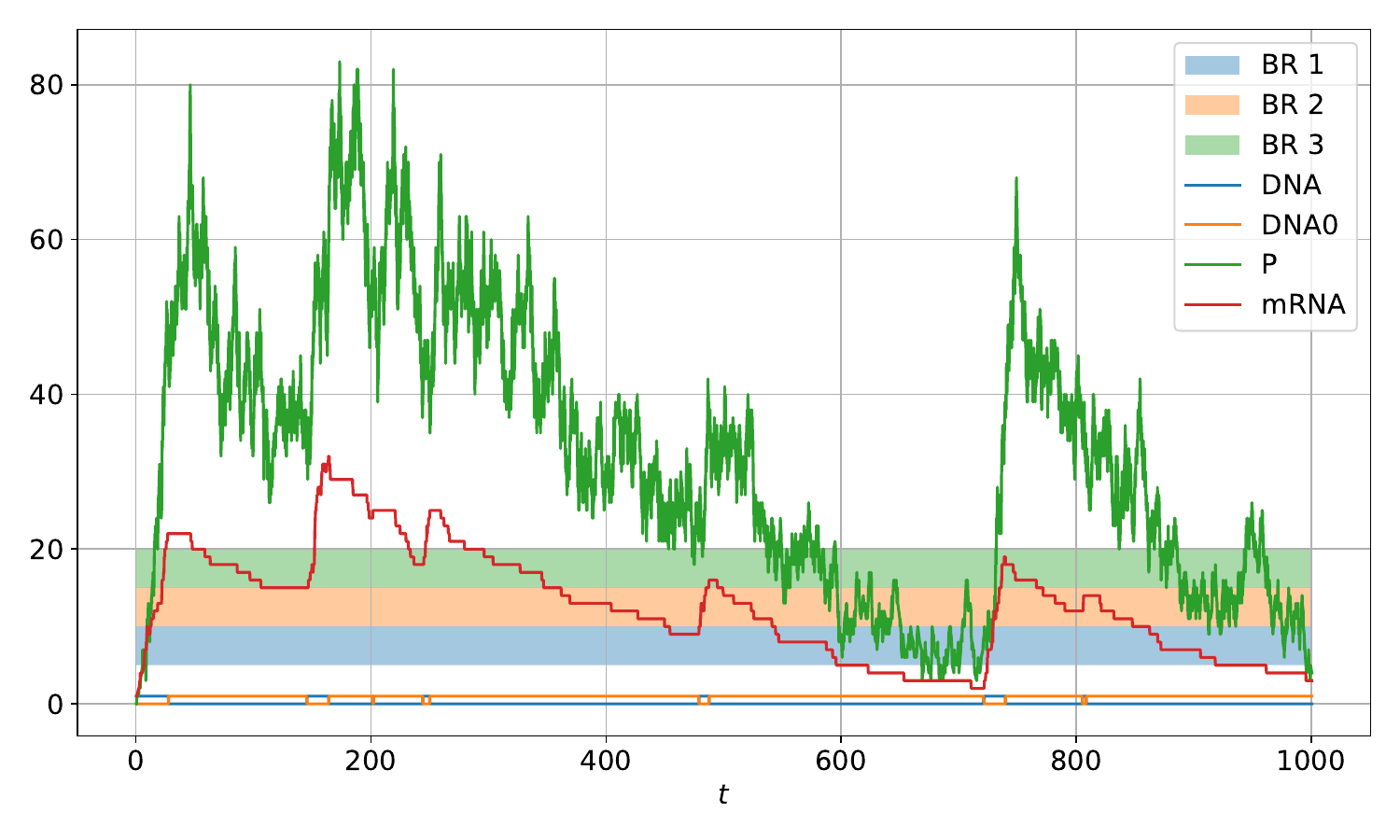}
    \caption{ \modtt{An SSA-simulated realisation of the Autorepressive Gene System. The system displays a multiple-scales structure. The mRNA and P species go through all possible simulation regime zones with \nmodtt{Blending Regions BR 1 = $(5,10)$, BR 2 = $(10, 15)$ and BR 3 = $(15,20)$.} }
    }
    \label{fig:ags-is-zones}
\end{figure}

\subsection{Parameter estimation}
\label{subsec:par}
This section investigates the reliability of the Hybrid $\tau$ algorithm for Bayesian inverse problems, in the setting described in Section~\ref{subsec:paramestim}

\subsubsection{Experiment setting}

The system on which the parameter estimation is performed is defined by the reactions (see also \cite{shgogi14}): 

\begin{system} \textbf{Bayesian inference system}
\begin{equation} \label{eq:bayinfsys}
    \begin{aligned}
        R_1 : \emptyset \xrightarrow{c_1} S_1 && R_3 : S_1 \xrightarrow{c_3} \emptyset && R_5 : S_1 + S_2 \xrightarrow{c_5} 20 S_2 \\
    R_2 : \emptyset \xrightarrow{c_2} S_2 && R_4 : S_2 \xrightarrow{c_4} \emptyset.
    \end{aligned} 
\end{equation}    
\end{system}
$R_1$ and $R_2$ are production reactions, $R_3$ and $R_4$ are decay reactions, and $R_5$ is a second-order reaction. The reaction rates are given by
\begin{equation*}
    \mathbf{c} = \left (2, sc, \sfrac{1}{50}, 1, \sfrac{1}{(50 \cdot  sc)} \right),
\end{equation*}
where $sc$ is used as a parameter, that takes the values~$sc~\in~\{1, 10, 10^2, 10^3\}$. We aim to infer the value of $\mathbf{c}$ for each of those values. To ensure identifiability, \modkz{$c_3$ is set to its true value $1/50$} in all simulations -- hence practically, the Bayesian inference is performed on $(c_1,c_2,c_4,c_5)$.  

Following~\cite{shgogi14}, the noisy measurements are assumed to follow the distribution
\begin{equation}
    \bY^n_i \knowingthat \bX_i(t_n) \sim \left\{
    \begin{aligned}
        & \mathrm{Poisson}(\bX_i(t_n)) && \mathrm{if} \; \modtt{\bX_i(t_n) > 0,}\\
        & \mathrm{Bernoulli}(0.1) && \mathrm{else}.
        \end{aligned}
    \right.
\end{equation}

\subsubsection{Methodology} \label{subsubsec:methodology}
\modk{We are now interested in calculating the posterior distribution \modtt{$\mathbb{P}(\bc|\by_{[0,T]})$}. We do this using the Algorithms  \ref{alg:ppmmh} and \ref{alg:bpf}.}

\modk{
There are a number of modeling choices and parameter tuning that need to be done before we proceed with our numerical experiments and comparisons. 
\modtt{
In this work, we use a Gaussian proposal kernel $q(\cdot \knowingthat \bc)$  with mean $\bc$ and a covariance $\gamma\hat{\mathbf{C}}$ which will be estimated using successice short exploratory runs. 
Using a first exploratory run of the particle filter with $\gamma=1, \hat{\mathbf{C}}=I$ (Algorithm \ref{alg:ppmmh}) allows us to determine a first estimate of the parameter by $\bc_{\mathrm{pre}}$ and the sample covariance matrix $\hat{\mathbf{C}}$. 
The number of particles $K$ is then chosen such that $\mathrm{Var}(\mathrm{log}\,\hat{\mathbb{P}}(\by\knowingthat \bc_{\mathrm{pre}}))$ lies between $0.25$ and $2.5$. Finally, the scaling parameter $\gamma$ is set such that the accept-reject ratio is close to $10\%$. \modtt{The choice of the number of particles $K$ and the acceptance probability are chosen following the guidelines outlined in~\cite{shthal15, pisigi12}.}}
}
\nmodtt{We summarise the obtained values in  Table \ref{tab:particle-filter-xp-params}.}

\begin{table}
\centering
\begin{tabular}{|l||ll|ll|}
\hline
            & \multicolumn{2}{c|}{\modkz{$K$}} & \multicolumn{2}{c|}{$\gamma$}   \\ \cline{2-5} 
 & \multicolumn{1}{c|}{SSA} & \multicolumn{1}{c|}{Hybrid $\tau$} & \multicolumn{1}{c|}{SSA} & \multicolumn{1}{c|}{Hybrid $\tau$} \\ \hline
$sc = 1$    & \multicolumn{1}{l|}{800}      & 320      & \multicolumn{1}{l|}{0.05} & 2.0 \\
$sc = 10$   & \multicolumn{1}{l|}{640}      & 1280     & \multicolumn{1}{l|}{4.0}  & 1.0 \\
$sc = 10^{2}$  & \multicolumn{1}{l|}{80}       & 160      & \multicolumn{1}{l|}{1.0}  & 1.0 \\
$sc = 10^{3}$ & \multicolumn{1}{l|}{320}      & 160      & \multicolumn{1}{l|}{2.0}  & 1.0 \\ \hline
\end{tabular}
\caption{Experimental values obtained for particle filter simulation.}
\label{tab:particle-filter-xp-params}
\end{table}

After those values are determined, a small burn-in phase of $200$ iterations is performed to ensure the particle filter starts in a credible region. 
Finally, the actual experiment is run with $n_{\mathrm{iter}} = 2 \cdot 10^{5}$ iterations, using the last explored state of the burn-in phase as the starting point. 
This procedure is repeated for each value of $sc$, and for simulations obtained via the SSA or via the Hybrid $\tau$.
\modtt{The tunable parameters in the Hybrid $\tau$ simulations were set to $I_1 = 20$, $I_2 = 60$ and $\Delta t = 10^{-1}$ and $\delta t = 10^{-2}$.}

The values $(N_{\mathrm{part}}, \gamma)$ chosen by in the setup phase boostrap are roughly of the same order for both forward simulators. This suggests that the Hybrid $\tau$ simulates the system in the same way as SSA, and thus causes the particle filter to explore the state space in the same way as (only much more efficiently when $sc$ becomes large). 


\subsubsection{Numerical Results}

\begin{figure}
   \centering
  \includegraphics[width=0.8\textwidth]{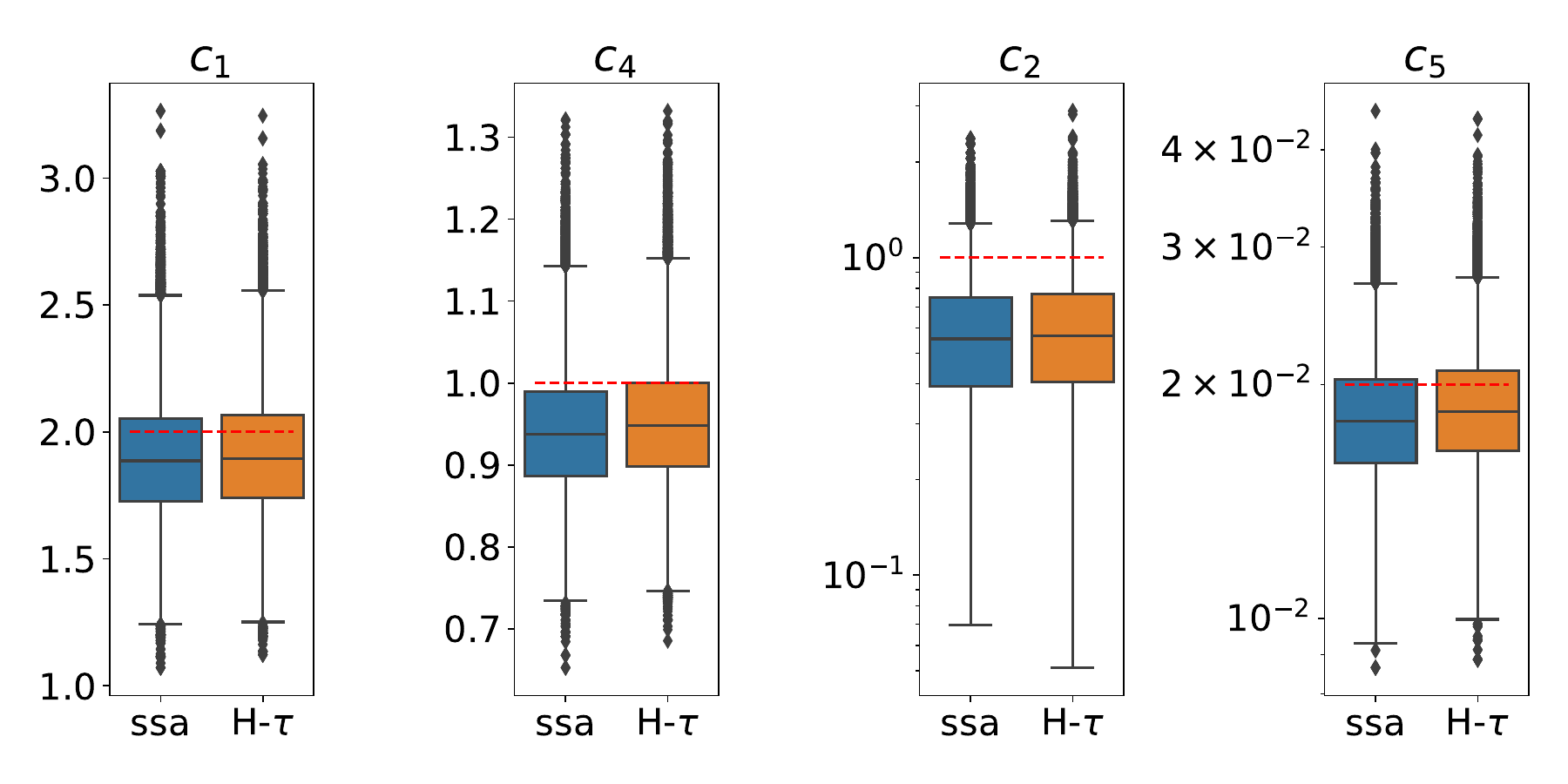}
     \caption{\nmodtt{Bayesian inference system. Box plots of samples of parameters $c_1, c_2, c_4, c_5$ from PPMMH algorithm, with simulator SSA and Hybrid $\tau$ respectively, for multiscale parameter $sc = 1$. The red line represents the true value to be inferred.}}
     \label{fig:vplot-1}
\end{figure}

\begin{figure}
     \centering
     \includegraphics[width=0.8\textwidth]{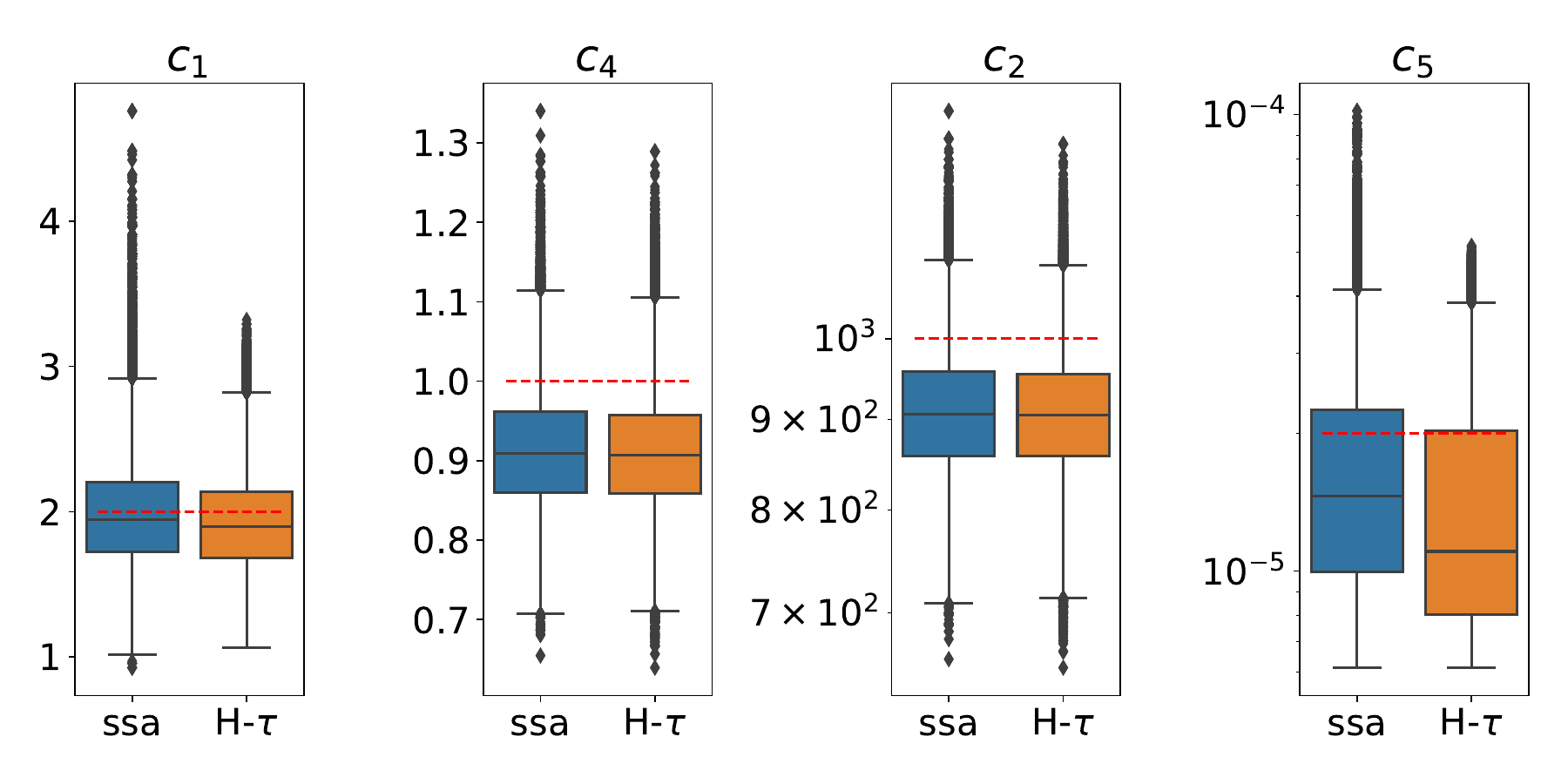}
     \caption{\nmodtt{Bayesian inference system. Box plots of samples of parameters $c_1, c_2, c_4, c_5$ from PPMMH algorithm, with simulator SSA and Hybrid $\tau$ respectively, for multiscale parameter $sc = 10^{3}$. The red line represents the true value to be inferred.}}
     \label{fig:vplot-1000}
\end{figure}

We run the inference procedure using either the standard SSA or the Hybrid $\tau$ algorithm. 
As can be seen in the box-and-whiskers plots (Figures~\ref{fig:vplot-1}--\ref{fig:vplot-1000}), across all scales the inference of the parameters $c_1,c_2,c_4,c_5$ are similar no matter which simulator is used. 
In all plots, the dashed red line represents the true value to be inferred. The colored boxes are delimited by the first and third quartiles, the whiskers have length $1.5 \times \mathrm{IQR}$ (interquartile range), and the points outside are statistical outliers. The values $c_1$ and $c_4$ have no multiscale dependence, while the parameters $c_2$ and $c_5$ do. 

\begin{figure}[h]
 \centering
  \includegraphics[width=0.7\textwidth]{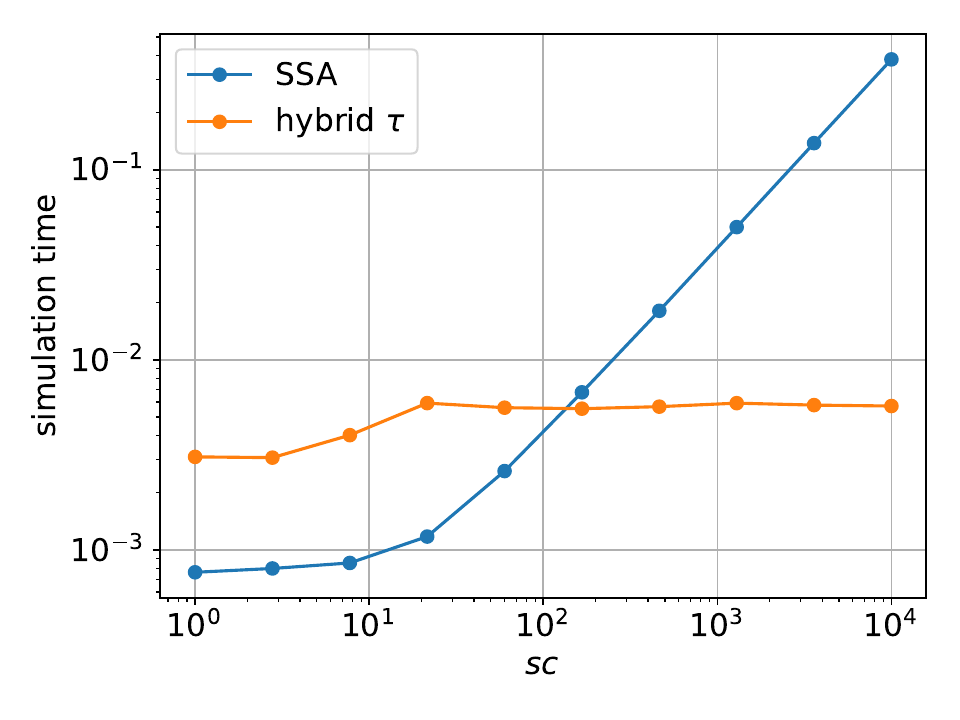}
    \caption{Mean simulation time for system~\eqref{eq:bayinfsys} (averaged over $100$ simulations) for different values of the parameter $sc$.}
   \label{fig:pf-times}
\end{figure}

The major difference is noted in the computational times, Figure~\ref{fig:pf-times} illustrates how, as the system becomes displays stronger multi-scale behaviour, the simulation time using the Hybrid $\tau$ algorithm is bounded, while the SSA simulation time grows linearly.

\section{Conclusions}
\label{sec:concl}

We have proposed a novel Hybrid $\tau$ algorithm, inspired by the Hybrid CLE method proposed in~\cite{zydunerb16}.
Unlike it, the quantities remain fully discrete at all times. The method leverages the splitting property of Poisson processes to rewrite the stochastic process as a sum of Poisson random variables that are simulated using SSA or $\tau$-leaping; the computational gains become obvious when one manages to ensure that the reactions fired frequently are simulated using the $\tau$-leaping, while those firing less often (and for which small stochastic variations may lead to important changes) are simulated using the exact SSA. 

We point out that this algorithm really belongs to a larger class of hybrid schemes based on the splitting of the Poisson process into separate processes that are simulated differently. In this sense, there is a significant amount of flexibility in the simulation of (multi-scale) stochastic chemical kinetics. Straightforwardly, one could swap the $\tau$-leaping by a more robust algorithm such as binomial $\tau$-leaping, $R$-leaping, etc. Other approaches such as a separation into multiple levels (e.g. micro-meso-macro) with a suitable simulation at each level are also possible. 

The choice of blending function discussed in this article is a ``cautious'' one, as it considers all the involved species in a given reaction (reactants and products). 
This can be understood as dynamically tagging a reaction to be simulated using the SSA if any species is low, $\tau$-leaping if all are abundant or a combination of both in the intermediate regime.  
Including the reactants in the blending functions makes intuitive sense, among others it can make the algorithm particularly robust to the issue of negative species. However, if it is known \textit{a priori} that certain products do not have a significant impact on system (no feedback loops, etc), not including them in the blending function may prove to be computationally interesting as it would naturally result in a larger state space in which the system is simulated with $\tau$-leaping.  

The various experiments presented in this article display how the Hybrid $\tau$ algorithm produces accurate and computationally efficient results. This is especially the case in the multi-scale setting. The algorithm is naturally amenable to that setting, as it handles it with no necessary additional modelling steps, unlike other algorithms specifically designed for the multi-scale case such as e.g. Piecewise Deterministic Markov processes in~\cite{winkelmann2017hmc}.   
The algorithm requires some user-prescribed values, the time-steps for the $\tau$-leaping and for the intermediate regime,
as well as the blending functions (additionally in our case, with their corresponding parameters $I_1$ and $I_2$). 
Numerically, it was observed that the dependence of the time-step sizes on the quality of the simulations is significantly more important than that of the blending regions. It would be of interest to investigate  \modk{further} the impact of the blending functions and regions, establishing criteria on how to suitably define them, as well as incorporating adaptive time-steps mechanisms for the different simulation regimes. \modkz{Additionally in the future, it would be interesting to adopt time-step strategies similar to the ones in \cite{motevi14} for choosing $\tau$. Finally, in terms of Bayesian inference an interesting research direction would be to see if the proposed framework could be modified  to allow particle trajectories to be conditioned on the next observation, similarly to the approach in \cite{GS19}, as this could allow for dealing with the shortcomings of the bootstrap particle filter. Similarly, it would be interesting to see if the proposed approach could be adjusted to perform inference using the procedure outlined in \cite{AAK23}. }

\section{\nmodtt{Appendix}}
We present here some tables summarising our numerical simulations for the autorepressive Gene-regulatory system. 
\begin{table}[h]
\centering
\begin{tabular}{|ccllllll|}
\hline
\multicolumn{1}{|l}{}     & \multicolumn{1}{l}{($\Delta t, \delta t$)}           & $I_1$ & $I_2$                   & DNA   & DNA$_0$ & Proteins & mRNA   \\ \hline
\multicolumn{1}{|c|}{}    & \multicolumn{1}{c|}{}                                & 5     & \multicolumn{1}{l|}{10} & 0.173 & 0.827   & 64.411   & 25.868 \\
\multicolumn{1}{|c|}{}    & \multicolumn{1}{c|}{}                                & 10    & \multicolumn{1}{l|}{15} & 0.134 & 0.866   & 63.041   & 25.267 \\
\multicolumn{1}{|c|}{}    & \multicolumn{1}{c|}{\multirow{-3}{*}{$(0.2, 0.15)$}} & 15    & \multicolumn{1}{l|}{20} & 0.111 & 0.889   & 61.763   & 24.734 \\ \cline{2-8} 
\multicolumn{1}{|c|}{}    & \multicolumn{1}{c|}{}                                & 5     & \multicolumn{1}{l|}{10} & 0.12  & 0.88    & 64.235   & 25.514 \\
\multicolumn{1}{|c|}{}    & \multicolumn{1}{c|}{}                                & 10    & \multicolumn{1}{l|}{15} & 0.151 & 0.849   & 63.974   & 25.401 \\
\multicolumn{1}{|c|}{\multirow{-6}{*}{H $\tau$}} & \multicolumn{1}{c|}{\multirow{-3}{*}{$(1, 0.4)$}} & 15 & \multicolumn{1}{l|}{20} & 0.134 & 0.866 & 63.286 & 25.54  \\ \hline
\multicolumn{1}{|c|}{}    & \multicolumn{1}{c|}{}                                & 5     & \multicolumn{1}{l|}{10} & 0.133 & 0.867   & 63.148   & 25.241 \\
\multicolumn{1}{|c|}{}    & \multicolumn{1}{c|}{}                                & 10    & \multicolumn{1}{l|}{15} & 0.125 & 0.875   & 62.039   & 24.962 \\
\multicolumn{1}{|c|}{}    & \multicolumn{1}{c|}{\multirow{-3}{*}{$(0.2, 0.15)$}} & 15    & \multicolumn{1}{l|}{20} & 0.121 & 0.879   & 63.462   & 25.546 \\ \cline{2-8} 
\multicolumn{1}{|c|}{}    & \multicolumn{1}{c|}{}                                & 5     & \multicolumn{1}{l|}{10} & 0.122 & 0.878   & 64.336   & 25.797 \\
\multicolumn{1}{|c|}{}    & \multicolumn{1}{c|}{}                                & 10    & \multicolumn{1}{l|}{15} & 0.133 & 0.867   & 61.483   & 24.616 \\
\multicolumn{1}{|c|}{\multirow{-6}{*}{H CLE}}    & \multicolumn{1}{c|}{\multirow{-3}{*}{$(1, 0.4)$}} & 15 & \multicolumn{1}{l|}{20} & 0.128 & 0.872 & 61.61  & 24.679 \\ \hline
\multicolumn{1}{|c|}{SSA} & \multicolumn{3}{l|}{\cellcolor[HTML]{C0C0C0}}                                          & 0.137 & 0.863   & 64.323   & 25.775 \\ \hline
\end{tabular}
\caption{Mean values of the Autorepressive Gene-regulatory System at $T=10^3$ with sample size $N = 10^4$ using SSA, Hybrid $\tau$ (H $\tau$) and Hybrid CLE with different parameters.}
\label{tab:ags-means}
\end{table}

\begin{table}[h!]
\centering
\begin{tabular}{|ccllllll|}
\hline
\multicolumn{1}{|l}{}     & \multicolumn{1}{l}{($\Delta t, \delta t$)}           & $I_1$ & $I_2$                   & DNA     & DNA$_0$ & Proteins & mRNA   \\ \hline
\multicolumn{1}{|c|}{}    & \multicolumn{1}{c|}{}                                & 5     & \multicolumn{1}{l|}{10} & 0.37825 & 0.37825 & 32.777   & 12.726 \\
\multicolumn{1}{|c|}{}    & \multicolumn{1}{c|}{}                                & 10    & \multicolumn{1}{l|}{15} & 0.34065 & 0.34065 & 32.699   & 12.789 \\
\multicolumn{1}{|c|}{}    & \multicolumn{1}{c|}{\multirow{-3}{*}{$(0.2, 0.15)$}} & 15    & \multicolumn{1}{l|}{20} & 0.31413 & 0.31413 & 33.77    & 13.286 \\ \cline{2-8} 
\multicolumn{1}{|c|}{}    & \multicolumn{1}{c|}{}                                & 5     & \multicolumn{1}{l|}{10} & 0.32496 & 0.32496 & 34.09    & 13.228 \\
\multicolumn{1}{|c|}{}    & \multicolumn{1}{c|}{}                                & 10    & \multicolumn{1}{l|}{15} & 0.35805 & 0.35805 & 34.61    & 13.492 \\
\multicolumn{1}{|c|}{\multirow{-6}{*}{H $\tau$}} & \multicolumn{1}{c|}{\multirow{-3}{*}{$(1, 0.4)$}} & 15 & \multicolumn{1}{l|}{20} & 0.34065 & 0.34065 & 33.797 & 13.271 \\ \hline
\multicolumn{1}{|c|}{}    & \multicolumn{1}{c|}{}                                & 5     & \multicolumn{1}{l|}{10} & 0.33957 & 0.33957 & 33.043   & 13.051 \\
\multicolumn{1}{|c|}{}    & \multicolumn{1}{c|}{}                                & 10    & \multicolumn{1}{l|}{15} & 0.33072 & 0.33072 & 32.518   & 12.759 \\
\multicolumn{1}{|c|}{}    & \multicolumn{1}{c|}{\multirow{-3}{*}{$(0.2, 0.15)$}} & 15    & \multicolumn{1}{l|}{20} & 0.32613 & 0.32613 & 32.486   & 12.933 \\ \cline{2-8} 
\multicolumn{1}{|c|}{}    & \multicolumn{1}{c|}{}                                & 5     & \multicolumn{1}{l|}{10} & 0.32729 & 0.32729 & 33.769   & 13.232 \\
\multicolumn{1}{|c|}{}    & \multicolumn{1}{c|}{}                                & 10    & \multicolumn{1}{l|}{15} & 0.33957 & 0.33957 & 33.414   & 13.085 \\
\multicolumn{1}{|c|}{\multirow{-6}{*}{H CLE}}    & \multicolumn{1}{c|}{\multirow{-3}{*}{$(1, 0.4)$}} & 15 & \multicolumn{1}{l|}{20} & 0.33409 & 0.33409 & 33.662 & 13.015 \\ \hline
\multicolumn{1}{|c|}{SSA} & \multicolumn{3}{l|}{\cellcolor[HTML]{C0C0C0}}                                          & 0.34385 & 0.34385 & 33.101   & 13.346 \\ \hline
\end{tabular}
\caption{Standard deviations of the Autorepressive Gene-regulatory System at $T=10^3$ with sample size $N = 10^4$ using SSA and Hybrid $\tau$ (H $\tau$) with different parameters.}
\label{tab:ags-stds}
\end{table}

\begin{table}[t]
\centering
\begin{tabular}{|llllll|}
\hline
                          & $(\Delta t, \delta t)$                               & $I_1$ & $I_2$                   & Simulation time & Ratio SSA \\ \hline
\multicolumn{1}{|l|}{}    & \multicolumn{1}{l|}{}                                & 5     & \multicolumn{1}{l|}{10} & 5.66759e-03     & 1.485     \\
\multicolumn{1}{|l|}{}    & \multicolumn{1}{l|}{}                                & 10    & \multicolumn{1}{l|}{15} & 5.92247e-03     & 1.421     \\
\multicolumn{1}{|l|}{}    & \multicolumn{1}{l|}{\multirow{-3}{*}{$(0.2, 0.15)$}} & 15    & \multicolumn{1}{l|}{20} & 6.47657e-03     & 1.3       \\ \cline{2-6} 
\multicolumn{1}{|l|}{}    & \multicolumn{1}{l|}{}                                & 5     & \multicolumn{1}{l|}{10} & 2.36346e-03     & 3.562     \\
\multicolumn{1}{|l|}{}    & \multicolumn{1}{l|}{}                                & 10    & \multicolumn{1}{l|}{15} & 2.72807e-03     & 3.086     \\
\multicolumn{1}{|l|}{\multirow{-6}{*}{H $\tau$}} & \multicolumn{1}{l|}{\multirow{-3}{*}{$(1, 0.4)$}} & 15 & \multicolumn{1}{l|}{20} & 3.46922e-03 & 2.426 \\ \hline
\multicolumn{1}{|l|}{}    & \multicolumn{1}{l|}{}                                & 5     & \multicolumn{1}{l|}{10} & 6.98658e-03     & 1.205     \\
\multicolumn{1}{|l|}{}    & \multicolumn{1}{l|}{}                                & 10    & \multicolumn{1}{l|}{15} & 7.31605e-03     & 1.151     \\
\multicolumn{1}{|l|}{}    & \multicolumn{1}{l|}{\multirow{-3}{*}{$(0.2, 0.15)$}} & 15    & \multicolumn{1}{l|}{20} & 8.07277e-03     & 1.043     \\ \cline{2-6} 
\multicolumn{1}{|l|}{}    & \multicolumn{1}{l|}{}                                & 5     & \multicolumn{1}{l|}{10} & 2.78733e-03     & 3.02      \\
\multicolumn{1}{|l|}{}    & \multicolumn{1}{l|}{}                                & 10    & \multicolumn{1}{l|}{15} & 3.28434e-03     & 2.563     \\
\multicolumn{1}{|l|}{\multirow{-6}{*}{H CLE}}    & \multicolumn{1}{l|}{\multirow{-3}{*}{$(1, 0.4)$}} & 15 & \multicolumn{1}{l|}{20} & 4.19815e-03 & 2.005 \\ \hline
\multicolumn{1}{|l|}{SSA} & \multicolumn{3}{l|}{\cellcolor[HTML]{C0C0C0}}                                          & 8.41800e-03     & 1         \\ \hline
\end{tabular}
\caption{Simulation times of the Autorepressive Gene-regulatory System at $T = 10^3$, averaged over $N=500$ realisations.}
\label{tab:ags-times}
\end{table}

\end{document}